\documentclass[a4paper,12pt]{article}
\pdfoutput=1 % if your are submitting a pdflatex (i.e. if you have)

\usepackage{jheppub_X} % for details on the use of the package, please
\usepackage{booktabs,colortbl}
\usepackage{cancel,xcolor,multirow}
\usepackage[normal]{caption}
\usepackage{subcaption}
\usepackage{slashed}
\usepackage{cleveref}
\usepackage{comment}
\usepackage[normalem]{ulem}

\crefname{table}{Table}{Tables}
\crefname{equation}{Eq.}{Eqs.}
\crefname{appendix}{App.}{Apps.}
\crefname{section}{Sec.}{Secs.}
\crefname{figure}{Fig.}{Figs.}

\usepackage[T1]{fontenc}
\usepackage{listings}
\usepackage[utf8]{inputenc}
\usepackage{lineno}
\newcommand{\tev}{~\text{TeV}}
\newcommand{\gev}{~\text{GeV}}
\newcommand{\nn}{\nonumber \\ }

\newcommand{\Lc}{\mathcal{L}}

\newcommand{\pythia}{\texttt{P}$_{\rm{NWA}}$}
\newcommand{\mg}{\texttt{MG}$_{\rm{Full}}$}
\newcommand{\ms}{\texttt{MS}}

\newcommand{\be}{\begin{equation}} 
\newcommand{\ee}{\end{equation}} 
\newcommand{\bea}{\begin{eqnarray}}  
\newcommand{\eea}{\end{eqnarray}}
\newcommand{\bs}{\begin{split}} 
\newcommand{\es}{\end{split}}

\newcommand{\pb}{~\mathrm{pb}}

\def\mnino{\ensuremath{m({\footnotesize \ninoone})}}
\def\mstop{\ensuremath{m({\footnotesize \stopone})}}
\def\mt{\ensuremath{m_t}}

 \def\stop{\ensuremath{\mathchoice%
      {\displaystyle\raise.0ex\hbox{$\displaystyle\tilde t$}}%
         {\textstyle\raise.0ex\hbox{$\textstyle\tilde t$}}%
       {\scriptstyle\raise.0ex\hbox{$\scriptstyle\tilde t$}}%
 {\scriptscriptstyle\raise.0ex\hbox{$\scriptscriptstyle\tilde t$}}}}
\def\stopone{\ensuremath{\mathchoice%
      {\displaystyle\raise.0ex\hbox{$\displaystyle\tilde t_1$}}%
         {\textstyle\raise-.2ex\hbox{$\textstyle\tilde t_1$}}%
       {\scriptstyle\raise.0ex\hbox{$\scriptstyle\tilde t_1$}}%
 {\scriptscriptstyle\raise.0ex\hbox{$\scriptscriptstyle\tilde t_1$}}}}
 
\def\nino{\ensuremath{\mathchoice%
      {\displaystyle\raise.4ex\hbox{$\displaystyle\tilde\chi^0$}}%
         {\textstyle\raise.4ex\hbox{$\textstyle\tilde\chi^0$}}%
       {\scriptstyle\raise.3ex\hbox{$\scriptstyle\tilde\chi^0$}}%
 {\scriptscriptstyle\raise.3ex\hbox{$\scriptscriptstyle\tilde\chi^0$}}}}
\def\ninoone{\ensuremath{\mathchoice%
      {\displaystyle\raise.4ex\hbox{$\displaystyle\tilde\chi^0_1$}}%
         {\textstyle\raise.1ex\hbox{$\textstyle\tilde\chi^0_1$}}%
       {\scriptstyle\raise.3ex\hbox{$\scriptstyle\tilde\chi^0_1$}}%
 {\scriptscriptstyle\raise.3ex\hbox{$\scriptscriptstyle\tilde\chi^0_1$}}}}

\lstdefinestyle{myCustomMatlabStyle}{
  language=Matlab,
  stepnumber=1,
  numbersep=10pt,
  tabsize=4,
  showspaces=false,
  showstringspaces=false
}

\definecolor{colorTC}{rgb}{.2,.7,.2}

\definecolor{colorSM}{rgb}{.7,.2,.7}
\definecolor{colorWH}{rgb}{.7,.2,.2}

\title{\Large 
Magnifying the ATLAS Stealth Stop Splinter: \\[5pt]
Impact of Spin Correlations and Finite Widths
}

\author{Timothy Cohen,}
\author{Walter Hopkins,}
\author{Stephanie Majewski,}
\author{and Bryan Ostdiek\\[-10pt]}

\affiliation{Institute of Theoretical Science and Center for High Energy Physics, \\[2pt] Department of Physics, University of Oregon, Eugene, Oregon 97403}

\abstract{
In this paper, we recast a ``stealth stop'' search in the notoriously difficult region of the stop-neutralino Simplified Model parameter space for which $\mstop - \mnino \simeq \mt$.  The properties of the final state are nearly identical for tops and stops, while the rate for stop pair production is $\mathcal{O}(10\%)$ of that for $t\bar{t}$.  Stop searches away from this stealth region have left behind a ``splinter'' of open parameter space when $\mstop \simeq \mt$.  Removing this splinter requires surgical precision:  the ATLAS constraint on stop pair production reinterpreted here treats the signal as a contaminant to the measurement of the top pair production cross section using data from $\sqrt{s} = 7 \text{ TeV}$ and $8 \text{ TeV}$ in a correlated way to control for some systematic errors.  ATLAS fixed $\mstop \simeq \mt$ and $\mnino = 1 \text{ GeV}$, implying that a careful recasting of these results into the full $\mstop - \mnino$ plane is warranted.  We find that the parameter space with $\mnino \lesssim 55 \text{ GeV}$ is excluded for $\mstop \simeq \mt$ --- although this search does cover new parameter space, it is unable to fully pull the splinter.  Along the way, we review a variety of interesting physical issues in detail: (\emph{i}) when the two-body width is a good approximation; (\emph{ii}) what the impact on the total rate from taking the narrow width is a good approximation; (\emph{iii}) how the production rate is affected when the wrong widths are used; (\emph{iv}) what role the spin correlations play in the limits.  In addition, we provide a guide to using \texttt{MadGraph} for implementing the full production including finite width and spin correlation effects, and we survey a variety of pitfalls one might encounter.}

\begin{document}
\maketitle
\flushbottom

%**************** Section ***********************************
\setcounter{page}{2}
\section{Introduction}
\label{sec:intro}
%*************************************************************
One of the main drivers of the Large Hadron Collider (LHC) program is to understand the origins of the electroweak scale.  In the Standard Model, the Higgs boson mass parameter $m_H^2$ is quadratically divergent, implying that it is not calculable within the Standard Model effective theory since it is sensitive to physics at arbitrarily high scales, \emph{e.g.}, the Planck scale where gravity becomes strong.  Fine-tuning away these quadratic contributions leads to the so-called hierarchy or naturalness problem.  One compelling way to ameliorate this problem is to assume that nature is Supersymmetric, thereby yielding a calculable Higgs boson mass parameter.  Then softly breaking this symmetry maintains the calculability of the theory, at the expense of introducing physical quadratic corrections to $m_H^2$.  Furthermore, every particle now gets a partner (whose spin differs by a half integer), implying a rich phenomenological program at the LHC to hunt down these new states. See~\cite{Martin:1997ns} for the classic review of Supersymmetry and~\cite{Giudice:2017pzm} for some perspective on the status of the naturalness problem.

While the masses of the superpartners are free parameters, the couplings are fixed by the symmetries of the theory.  This implies that, given limits on a mass spectrum, one can compute the fine-tuning required to reproduce the measured electroweak scale.  Unsurprisingly, the two superpartners with the largest couplings play the most important role: the superpartners of the top -- the stop -- and the gluon -- the gluino.  The mass of the partner of the Higgs is also relevant, but the limits on its mass are sufficiently weak due to a low production cross section such that it is not as relevant as the stop and gluino.  Taking the connection between the spectrum and the fine-tuning problem seriously motivates that the stop, gluino, and Higgsinos are the lightest new physics states that would appear at the LHC~\cite{Dimopoulos:1995mi, Cohen:1996vb}.  This spectrum has become even more compelling in the light of the comprehensive set of constraints on new physics published by both ATLAS and CMS, as emphasized in~\cite{Papucci:2011wy, Brust:2011tb, Buckley:2016tbs}.  This has additionally motivated many studies recasting ATLAS and CMS results to comprehensively understand the connection between the allowed parameter space for Higgsinos, stops, and gluinos, and the required tuning, \emph{e.g.}~\cite{Evans:2013jna, Papucci:2014rja, Buckley:2016kvr}.  

\begin{figure}[t!]
\begin{center}
%\vspace{-2cm}
\includegraphics[width=\linewidth]{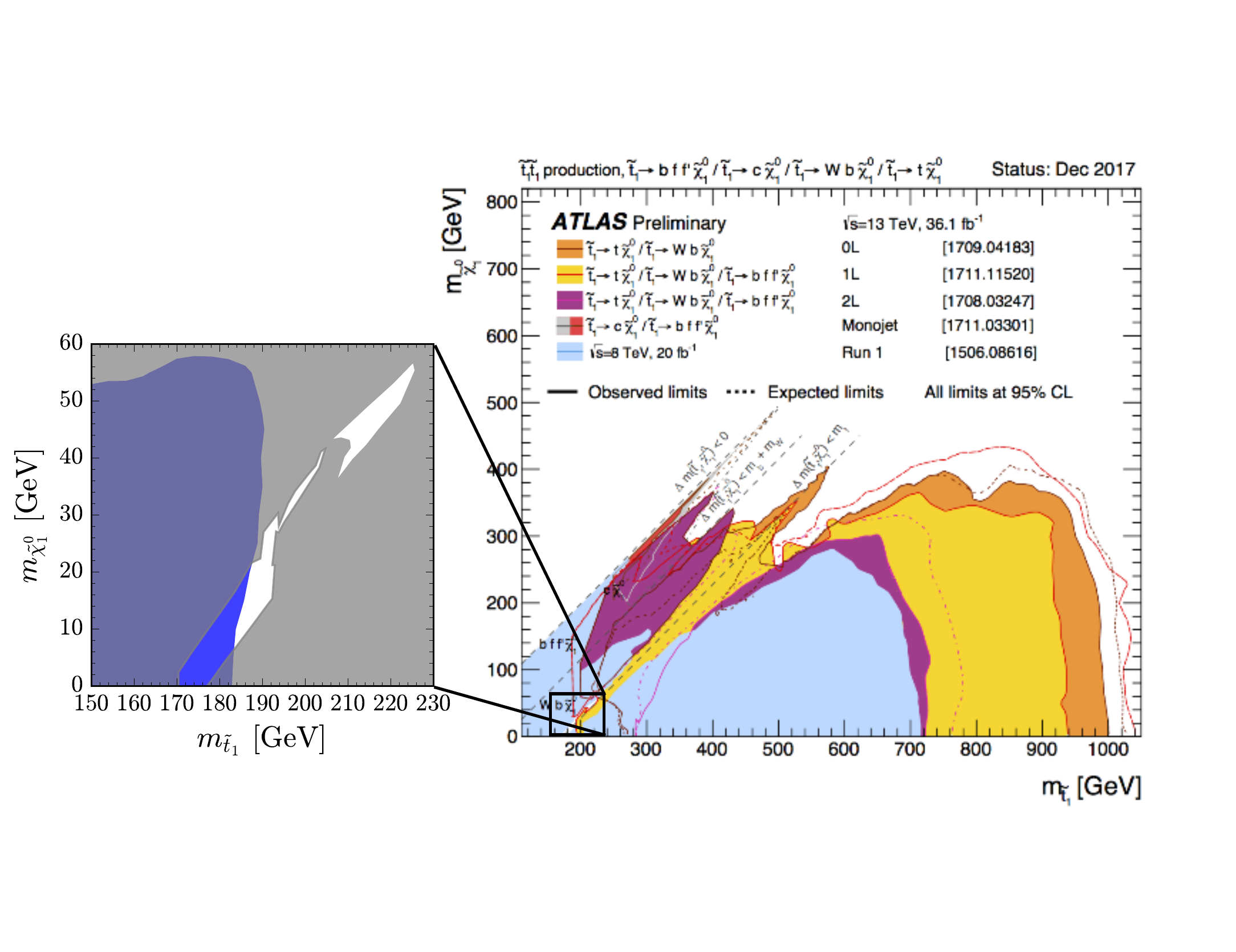}
%\vspace{-2cm}
\caption{The blue region in the breakout box shows the results of our recast. The gray region summarizes the ATLAS constraints, not including the $t\bar{t}$ spin correlation constraint, or the $\sigma_{t\bar{t}}$ contamination constraint (which is the subject of this paper)~\cite{SUSYSummaryPlot}.}
\label{fig:Exclusion}
\end{center}
\end{figure}

The focus of the paper here will be on the stop-neutralino ($\stopone$-$\ninoone$) Simplified Model, with a focus on the status of the parameter space currently constrained by ATLAS.  In particular, we want to explore what is still allowed in the very difficult ``stealth stop'' region where $\mstop -\mnino \simeq m_t$, where $m_t$ is the top quark mass.  The compressed kinematics make this particularly difficult.  Although the final state is $t\bar{t}\,\ninoone\,\ninoone$, the neutralinos are nearly at rest so that the resulting missing energy is very small and thus the observable signature is a pair of top quarks.  Furthermore, the production cross section for stop pair production is $\mathcal{O}(10\%)$ of the top pair production cross section.  Given these challenges, ATLAS has done a remarkable job constraining the stealth parameter space, and at this point only a ``splinter'' remains~\cite{SUSYSummaryPlot} as shown in the gray region of \cref{fig:Exclusion}.  

In more detail, ATLAS constraints on stealth stop production are derived through several independent analysis approaches: searches for direct stop pair production where $\mstop \gtrsim m_t$, $\stopone \rightarrow t + \ninoone$ and the top decay is on-shell~\cite{SUSY-2016-15,SUSY-2016-16}; searches for direct stop pair production where $\mstop -\mnino \lesssim m_t$ and the stops each decay via $\stopone \rightarrow W\, b\, \ninoone$~\cite{SUSY-2016-17}; analysis of $t\bar{t}$ spin correlations through an angular analysis of the lepton decay products~\cite{TOPQ-2014-07}; and a simultaneous measurement of the $t\bar{t}$ production cross section at $\sqrt{s}=7 \tev$ and $8 \tev$~\cite{TOPQ-2013-04}. The latter two analyses constrain stop pair production in the stealth region through detailed measurements of $t\bar{t}$ processes. CMS has performed similar searches, \emph{e.g.}~\cite{Sirunyan:2017cwe,Sirunyan:2017kqq,Sirunyan:2017fsj,Sirunyan:2017mrs,Sirunyan:2017uyt,Sirunyan:2017hvp}. There have additionally been some relevant phenomenology studies providing complementary ideas for targeting stealth stops~\cite{Han:2012fw, Alves:2012ft, Eifert:2014kea, Czakon:2014fka, Buckley:2014fqa, Fuks:2014lva, Ferretti:2015dea, Kobakhidze:2015scd}.

Our focus here will be to recast the limit derived from $\sigma_{t\bar{t}}$~\cite{TOPQ-2013-04} into the full $\mnino$-$\mstop$ plane --- ATLAS only provides results for fixed $\mnino = 1 \gev$.  A careful theoretical treatment is needed in order to properly simulate events in the region where $\mstop -\mnino \simeq m_t$.  In particular, a naive factorizing of the production and decay for the stop production misses important physical effects from the finite size of the widths involved, and from spin correlations.  Our goal here is to review these issues in detail, with an emphasis on their impact for recasting the results of~\cite{TOPQ-2013-04}.  We also provide a guide for the reader for simulating events near threshold using modern Monte Carlo event generation tools \texttt{MadGraph} and \texttt{Pythia}.  These lessons are more generic than the application presented here, and our hope is that the reader will find them useful when simulating events in a region where mass thresholds are relevant.

This paper is organized as follows.  A review of the theoretical subtleties that arise when simulating stealth stop production/decay (or more generally for generating events near a mass threshold) through the splinter is presented in \cref{sec:calculation}; additional technical details can be found in \cref{sec:eventgen} and some common pitfalls one might encounter are discussed in \cref{sec:pitfalls}.  Then we provide a review of the ATLAS top cross section measurement, and stealth stop limit in \cref{sec:search}.  Emphasis is given to the CL$_s$ limit setting procedure utilized here, and a validation of our framework is presented.   Finally, \cref{sec:recast} provides the results of our recast and shows how much of the splinter can be excluded.  We further emphasize the role of the systematic error on the top cross section by showing how the limits can be strengthened if this is reduced.   A brief closing discussion is given in~\cref{sec:discussion}.

%**************** Section ***********************************
\section{Calculating Stealth Stop Pair Production}
\label{sec:calculation}
%*************************************************************

A self-consistent framework for generating events is required to properly reinterpret the experimental constraints in the stealth stop region where $\mstop - \mnino \simeq \mt$. We address four effects that could impact the recasted limit: carefully computing the decay width in the two-to-three-body transition region, the error introduced by the narrow width approximation (NWA), the importance of including width in the Breit-Wigner propagator, and how spin correlations can change the efficiency to pass the signal region cuts. These are each detailed in Secs.~\ref{sec:2v3} to \ref{sec:SpinCorr}.

Three approaches are compared.  All three use \texttt{MadGraph5\_aMC@NLO 2.6.1}~\cite{Alwall:2014hca} for the hard matrix element calculation and convolution with the \texttt{NNPDF3.0} leading order parton distribution functions~\cite{Ball:2014uwa}, and \texttt{Pythia 8.2}~\cite{Sjostrand:2014zea} is always used for the parton shower and hadronization.  The stop pair production cross section is calculated using \texttt{NLLFast}~\cite{Beenakker:2015rna, Beenakker:1996ed, Beenakker:1997ut}; we additionally correct for the numerical impact of the NWA (see \cref{sec:NWA}) when using the \texttt{NLLFast} cross section to compute our final constraints.  Detector effects are approximated using \texttt{Delphes 3.4.1}~\cite{deFavereau:2013fsa}. We use the default \texttt{delphes\_card\_ATLAS.tcl} card, modified so that the jet radius is 0.4, in accordance with ATLAS. Jets are clustered with the anti-$k_t$ algorithm~\cite{Cacciari:2005hq,Cacciari:2008gp} within the \texttt{FastJet 3.2.1}~\cite{Cacciari:2011ma} framework.   More information regarding the event generation is given in \cref{sec:eventgen}. 

\begin{figure}[t]
\begin{center}
\includegraphics[width=0.45\linewidth]{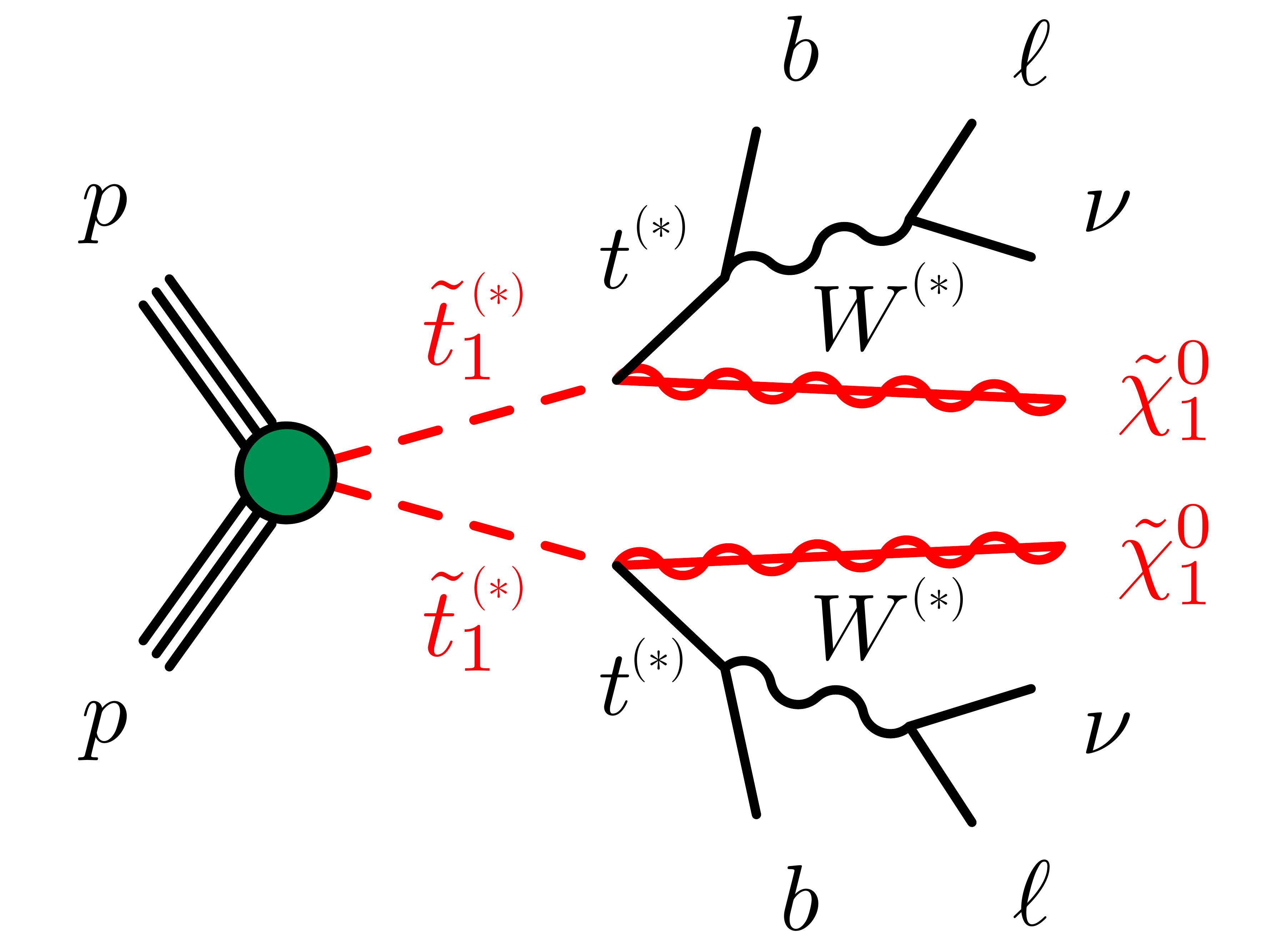}%
\hspace{30pt}
\includegraphics[width=0.45\linewidth]{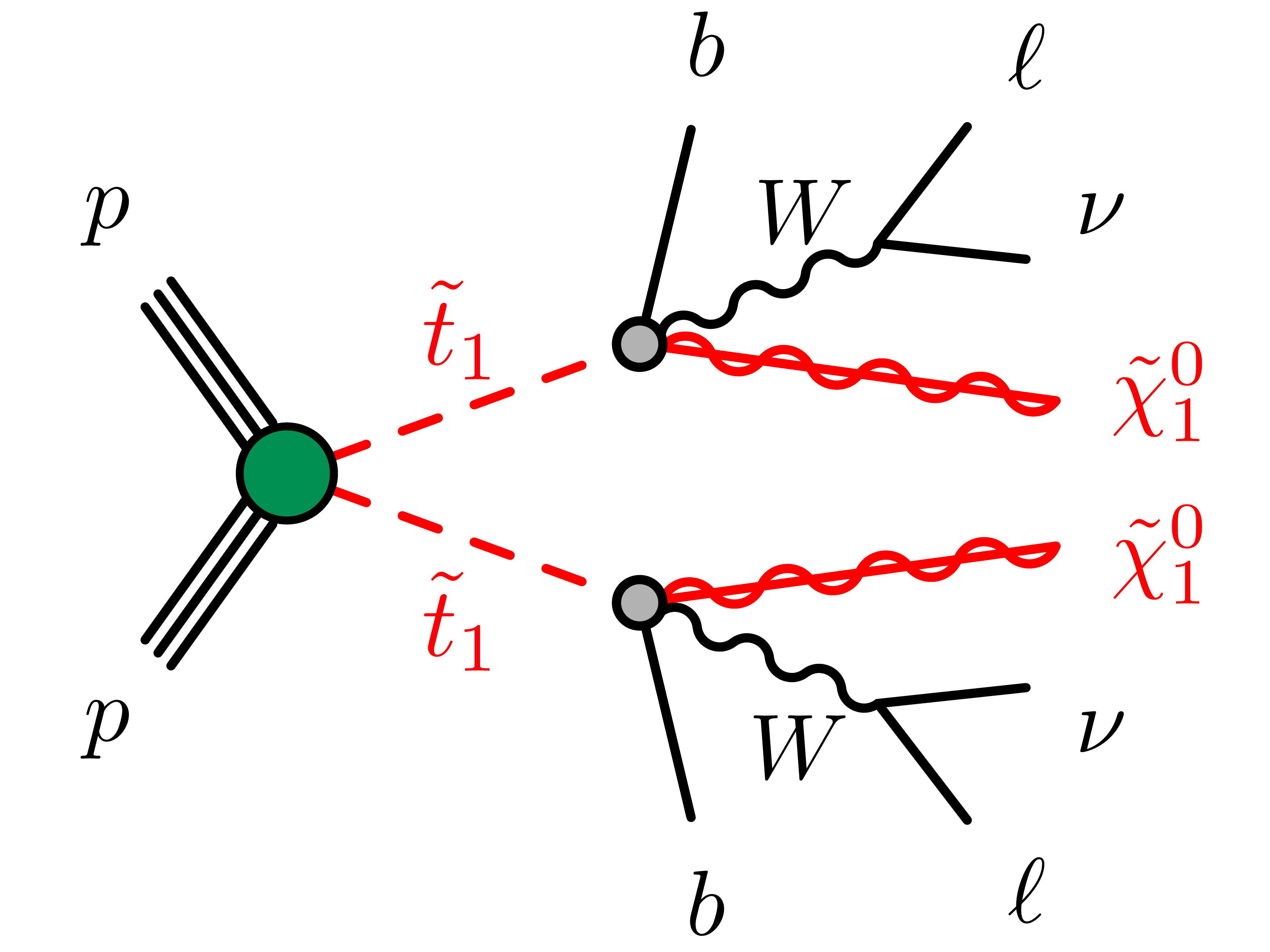}%
\caption{The left diagram illustrates the \mg\ scheme and the right diagram illustrates the \pythia\ scheme for the parameter space where $\mstop -\mnino < m_t$ (see \cref{tab:EvGenComparison} for an explanation of the naming conventions).  The green circles represent the full tree-level stop pair production matrix element.  The gray circles represent decays that do not include any matrix element information, \emph{i.e.}, the particles are decayed using phase space alone.  The superscripts $(*)$ denote particles that can go off shell.  This figure was adapted from the diagrams in~\cite{SUSY-2016-17}.}
\label{fig:FeynmanDiagrams}
\end{center}
\end{figure}

The differences come into what approximation is assumed when decaying the stops.  The relevant Feynman diagrams for stop production\footnote{Here we will follow the ATLAS conventions and express stop pair production as $p\,p \rightarrow \stopone\,\stopone$ instead of $p\,p \rightarrow \stopone\, \stopone^*$.} and decay are given in \cref{fig:FeynmanDiagrams}.   In both the left and right diagrams, the circle connecting the proton lines represents all of the possible diagrams that contribute to stop pair production, assuming the only particles beyond the Standard Model are the lightest stop and lightest neutralino.   In the left diagram, the stops decay to a neutralino and a top quark, which can be off shell. The top quark then decays to a bottom quark and a $W$ boson, which further decays leptonically (so as to populate the signal region).  This is the diagram for the most complete calculation (which we use for computing the recasted limits below), implemented using \texttt{MadGraph} to compute the full parton-level final state (before showering), and is referred to as \mg.  It can also be interpreted as the calculation performed by \texttt{MadSpin}~\cite{Artoisenet:2012st} (the approach taken by ATLAS), which computes spin correlations taking all intermediate particles exactly on-shell; we refer to this approach as \ms.  The final approach (\pythia) is to allow \texttt{Pythia} to decay the stops, which is equivalent to assuming a flat matrix element \emph{i.e.}~neglecting any spin effects in the final state.  This is illustrated by the diagram on the right, where the stop squark goes through a three-body decay directly.  The naming conventions for these three approaches are summarized in \cref{tab:EvGenComparison}. 
 
\begin{table}[h!]
\renewcommand{\arraystretch}{1.7}
\setlength{\tabcolsep}{10pt}
\setlength{\arrayrulewidth}{2pt}
%\begin{tabular}{l !{\vrule width 1pt} c c c c c c c c c c}
\begin{center}
   \begin{tabular}{c !{\vrule width 1.5 pt} c !{\vrule width 1.5 pt} c !{\vrule width 1.5 pt} c} % Column formatting, @{} suppresses leading/trailing space
   	Abbreviation & Matrix Element & Stop Decay & Shower/Hadron \\ \hline
	\mg & \texttt{MadGraph} & \texttt{MadGraph} & \texttt{Pythia} \\
	\ms & \texttt{MadGraph} & \texttt{MadSpin} & \texttt{Pythia} \\
	\pythia & \texttt{MadGraph} & \texttt{Pythia} & \texttt{Pythia} \\
   \end{tabular}
  \caption{The naming conventions for the various event generation tools used. }
   \label{tab:EvGenComparison}
\end{center}
\end{table}
 
There are many subtleties that must be accounted for to correctly evaluate production for the parameter space near the three- to two-body threshold.  Three-body effects on the width can be sizable, even if the two-body final state is open.  The narrow width approximation yields percent level errors with regard to the full production cross sections and/or width calculations.  If one uses the wrong width, this can have a non-trivial impact on the total rate.  Finally, spin correlations affect the distributions of the final state particles, and are particularly important when the three-body decay dominates.  The next four subsections explain these points in detail.

%**************** Section ***********************************
%\clearpage
\subsection{Transition from Three-body to Two-body Decays}
\label{sec:2v3}
%*************************************************************

If the channel is open, two-body decays nearly always dominate three-body decays due to phase space suppression. However, this transition must be accounted for continuously, which requires that care must be taken near threshold.

\begin{figure}[b]
\begin{center}
\includegraphics[width=0.5\linewidth]{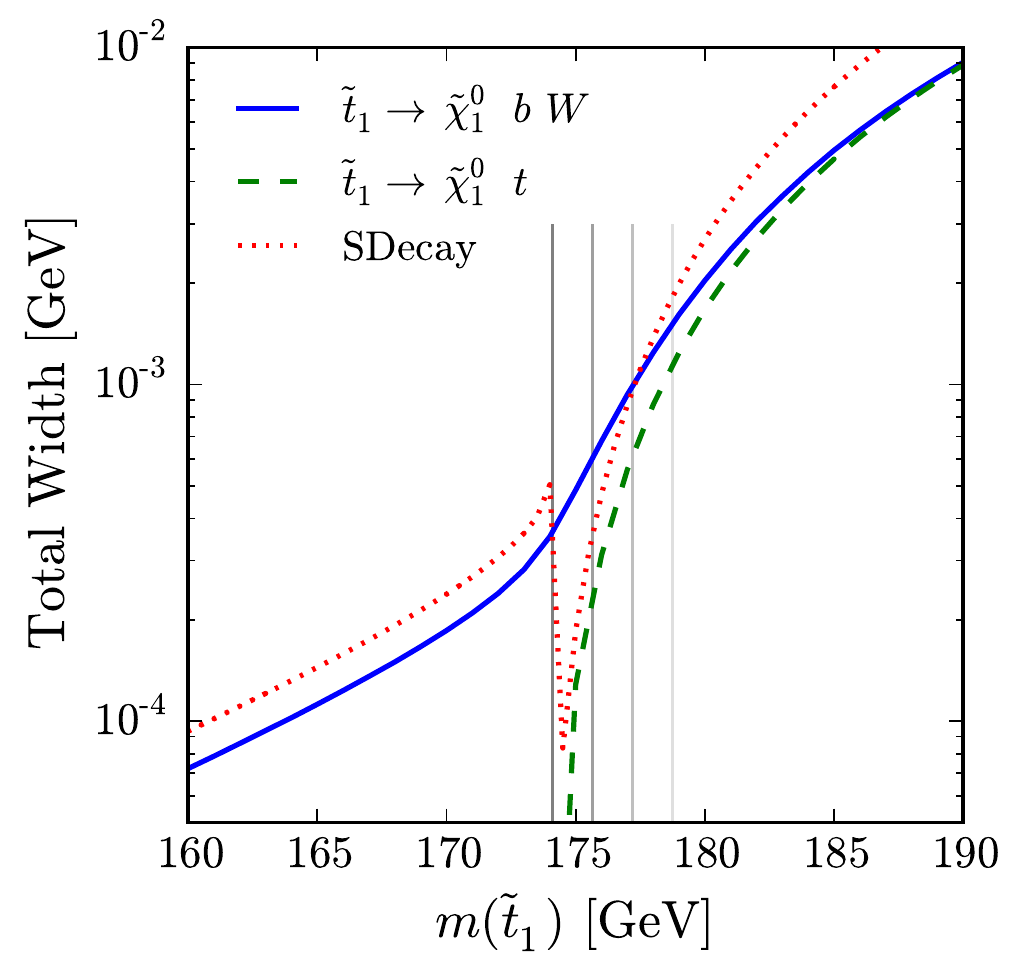}
\caption{The total width of a right handed stop computed assuming $m_t$ = 173.1 GeV and $\mnino =1\gev$.  The solid blue and dashed green lines are computed with \texttt{MadGraph}, by forcing a two- and three-body decay respectively. Clearly, the two-body approach does not capture near threshold effects (which can be important with multiple widths of the top as denoted by the thin vertical lines), due to the phase space factor implicit in the two-body assumption. For reference, we also show a comparison with \texttt{SDecay} in the red dot-dashed line, since it includes loop corrections from QCD. However, clearly \texttt{SDecay} is not reliable near threshold.}
\label{fig:DecayWidths}
\end{center}
\end{figure}

In \cref{fig:DecayWidths}, the stop decay width is plotted as a function of the stop mass. The blue and dashed, green lines are computed with \texttt{MadGraph}, and are the two- or three-body decays, respectively. Well above threshold, the two- and three-body decays yield the same width, which implies that the three-body process is dominated by the on-shell (as opposed to an off-shell) top quark. However, as the mass of the stop is lowered towards the threshold, the two-body processes start to diverge as there is less available phase space for the two-body decay.  The darkest vertical line denotes the threshold, while the lighter ones mark one, two, and three top widths above threshold.  We see that the stop mass needs to be many top widths above threshold for the off-shell top to not contribute significantly.

While \cref{fig:DecayWidths} shows that off-shell effects are important even above threshold, not all tools will carefully take this into account since their purpose is usually to cover the parameter space where there are no accidental degeneracies. For instance, the red line displays the width of the stops as calculated by \texttt{SDecay}~\cite{Muhlleitner:2003vg}. As soon as the stop is above threshold, the program only computes the two body diagram, leading to a discontinuity in the stop width. Similarly, using the default \texttt{compute\_widths} command during event generation in \texttt{MadGraph} also only computes two-body modes anywhere above threshold. The offset between the \texttt{SDecay} and \texttt{MadGraph} results is due to the inclusion of QCD corrections to the width by \texttt{SDecay}. The impact of incorrectly computing the stop width is explored further in \cref{sec:WrongWidths} and \cref{sec:pitfalls}.

%**************** Section ***********************************
%\clearpage
\subsection{Narrow Width Approximation}
\label{sec:NWA}
%*************************************************************

The Narrow Width Approximation (NWA)~\cite{Veltman:1963th, Dicus:1984fu} is typically assumed when intermediate unstable particles can contribute on-shell to processes of interest. We review the approximation here; see \emph{e.g.}~\cite{Veltman:1963th,Dicus:1984fu,Berdine:2007uv,Kauer:2007zc,Uhlemann:2008pm} for additional discussions.

The essential physics is straightforward to understand.  An unstable particle with mass $M$ and width $\Gamma$ contributes to a process $\mathcal{P}$ via a propagator:
\begin{equation}
\mathcal{P} = \int_{-\infty}^\infty \frac{\text{d}q^2}{2\,\pi}\left|\frac{1}{q^2 - M^2 + i\,\Gamma\,M} \right|^2\, \widetilde{\mathcal{M}}(q^2)\, ,
\end{equation}
where $q$ is the momentum flowing through the propagator of interest, and $\widetilde{\mathcal{M}}(q^2)$ represents the rest of the (integrated and spin averaged) matrix element squared for $\mathcal{P}$. Note that $\mathcal{P}$ captures both decay and production. When the width is small compared to the mass of the particle, the propagator is highly peaked near $q^2 \sim M^2$. The rest of the matrix element can then be evaluated at $q^2 = M^2$ using the integrated propagator:
\begin{equation}
\int_{-\infty}^\infty \text{d}q^2\,\frac{1}{(q^2 - M^2)^2 + \Gamma^2\, M^2}  = \frac{\pi}{M\,\Gamma}\,.
\end{equation}
Thus in the NWA, $\mathcal{P}$ is given by:
\begin{equation}
\mathcal{P}  
 \simeq \frac{1}{2\,M\,\Gamma} \,\widetilde{\mathcal{M}}(M^2) \, .
\label{eq:NWAeq}
\end{equation}

As a toy example with which we can investigate the quality of the NWA, we compute scalar production through an $s$-channel propagator convolved with the gluon pdf.\footnote{This toy model can be seen as $\phi\,\phi \rightarrow \Phi \rightarrow \phi\, \phi$ production, where $\phi$ is a massless scalar whose initial state momentum distribution is assumed to follow the gluon pdf, $\Phi$ is a massive scalar in the $s$-channel, and they interact via a three-point coupling $\mathcal{L} \supset a\,\Phi\,\phi^2$ for some non-zero coupling $a$.}  The production is given by the full propagator:
\begin{equation}
\mathcal{P}_{\rm{Full}} = A \int \text{d}Y \int \text{d}q^2 ~ \frac{q^2}{s}\,f_g \bigg(\frac{q}{\sqrt{s}}\, e^Y\bigg) f_g \bigg(\frac{q}{\sqrt{s}}\, e^{-Y} \bigg) \frac{1}{(q^2 - M^2)^2 + \Gamma^2 M^2}\, ,
\label{eqn:procFull}
\end{equation}
where $A$ is a constant that includes the couplings and phase space factors, $Y$ is the rapidity of the final state particles, and $\sqrt{s}$ is the center of mass energy of the protons.  For comparison, the process in the NWA is:
\begin{equation}
\mathcal{P}_{\rm{NWA}} = A \int \text{d}Y  ~ \frac{M^2}{s}\,f_g \bigg(\frac{M}{\sqrt{s}}\, e^Y\bigg) f_g \bigg(\frac{M}{\sqrt{s}}\, e^{-Y} \bigg) \frac{\pi}{M\, \Gamma} \,.
\label{eqn:procNWA}
\end{equation}
Eqs.~(\ref{eqn:procFull}) and (\ref{eqn:procNWA}) are evaluated using the \texttt{MSTW 2008} NNLO parton distributions~\cite{Martin:2009iq, Martin:2009bu, Martin:2010db} at a center-of-mass energy of $\sqrt{s}=8$ TeV for a variety of masses and widths.

\begin{figure}[t]
\begin{center}
\includegraphics[width=0.95\linewidth]{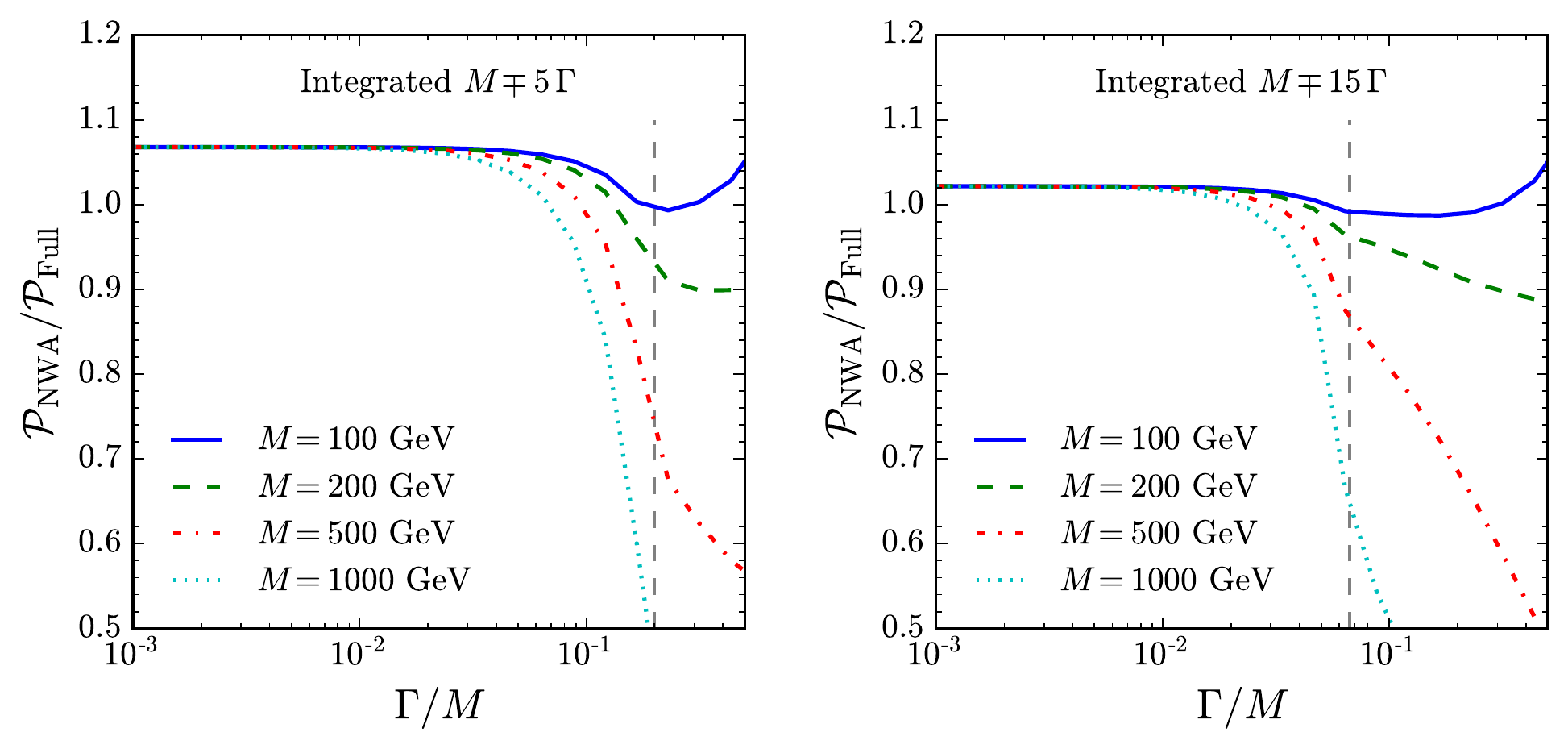}
\caption{This figure shows a comparison between integrating over a full squared matrix element or using the narrow width approximation by plotting $\mathcal{P}_\text{NWA}/\mathcal{P}_\text{Full}$. The left and right panels show the result when the integration limits are 5 or 15 widths around the resonance peak, respectively. The vertical dashed lines show the value of the width-to-mass ratio which would lead to unphysical imaginary momenta in the integration; values to the right use a constant lower bound for the integration instead.  We see that the NWA is no longer a good approximation for $\Gamma / M\gtrsim 10^{-2}$, and that this breakdown happens earlier as the mass is increased.}
\label{fig:NWAimpact}
\end{center}
\end{figure}

As the gluon distribution grows at low momentum fractions, we cannot perform the momentum integral in \cref{eqn:procFull} over the full phase space. To mitigate this issue, we fix the limits of integration in terms of a given number of widths around the resonance peak. Figure~\ref{fig:NWAimpact} shows the ratio of the NWA of \cref{eqn:procNWA} to the full calculation of \cref{eqn:procFull} as a function of the width-to-mass ratio $\Gamma/M$.  The integration range is taken to be 5 or 15 widths in the left and right panels, respectively, and the different colors show different choices of resonant mass.  Due to the choice for fixing the limits of integration in terms of a number of widths, there is an intrinsic limit to how large the width-to-mass ratio can be; if the width is too large, the integral range implies imaginary momenta.  When this occurs (as marked by the vertical dashed lines in the figure), a constant starting lower bound for the integration is used. 

For the simple toy process modeled here, at small widths the NWA overestimates the rate by 6\% or 2\% for integration windows of 5 or 15 widths, respectively. The NWA is no longer effective when the width is a few percent of the mass, and larger masses imply a faster breakdown.  This seemingly small effect can be important as the NWA is commonly used, and this discrepancy factor naively contributes for each NWA assumed in a diagram. For instance, producing the particles on-shell with subsequent decays left to other programs implicitly assumes the NWA as it does not integrate over the intermediate on-shell propagator. As shown in the appendix, this yields a 6\% effect on the total cross section for stop pair production. The NWA is also usually assumed when there are particle decay chains, such as when computing the width of the top quark -- integrating over the intermediate $W$ from a top decay leads to a few percent effect.

%**************** Section ***********************************
%\clearpage
\subsection{Inconsistent widths}
\label{sec:WrongWidths}
%*************************************************************
As shown above, common practice approaches to width calculations near threshold can yield errors near a factor of 10. In this subsection, we examine the numerical impact on the production rate if one uses the wrong width. We will specialize to the stop pair production and decay process of interest:
\begin{align}
p\,p \rightarrow \stopone\,\stopone \rightarrow b\,\bar{b}\,f\,\bar{f}'\, f''\, \bar{f}'''\,\ninoone\,\ninoone\,,
\end{align}
where we do not put any kinematic requirements on the internal top or $W$.  

By inspection of \cref{fig:FeynmanDiagrams}, two stop propagators must be considered.  Since these widths are very small, we can reliably estimate how an error in the stop width propagates to the full process using the NWA in \cref{eq:NWAeq}. If the stop width is computed inconsistently by an amount $\Delta\Gamma$, then including the effect of both stop propagators implies that the cross section prediction will be wrong by approximately
\begin{equation}
\sigma_{\Delta\Gamma} \simeq \left(\frac{\Gamma}{\Gamma+\Delta\Gamma}\right)^2 \sigma \simeq \left(1- 2\,\frac{\Delta\Gamma}{\Gamma} + \,\cdots \right) \sigma\,.
\label{eqn:WidthErrorEstimate}
\end{equation}

\begin{figure}[t]
\begin{center}
\includegraphics[width=0.95\linewidth]{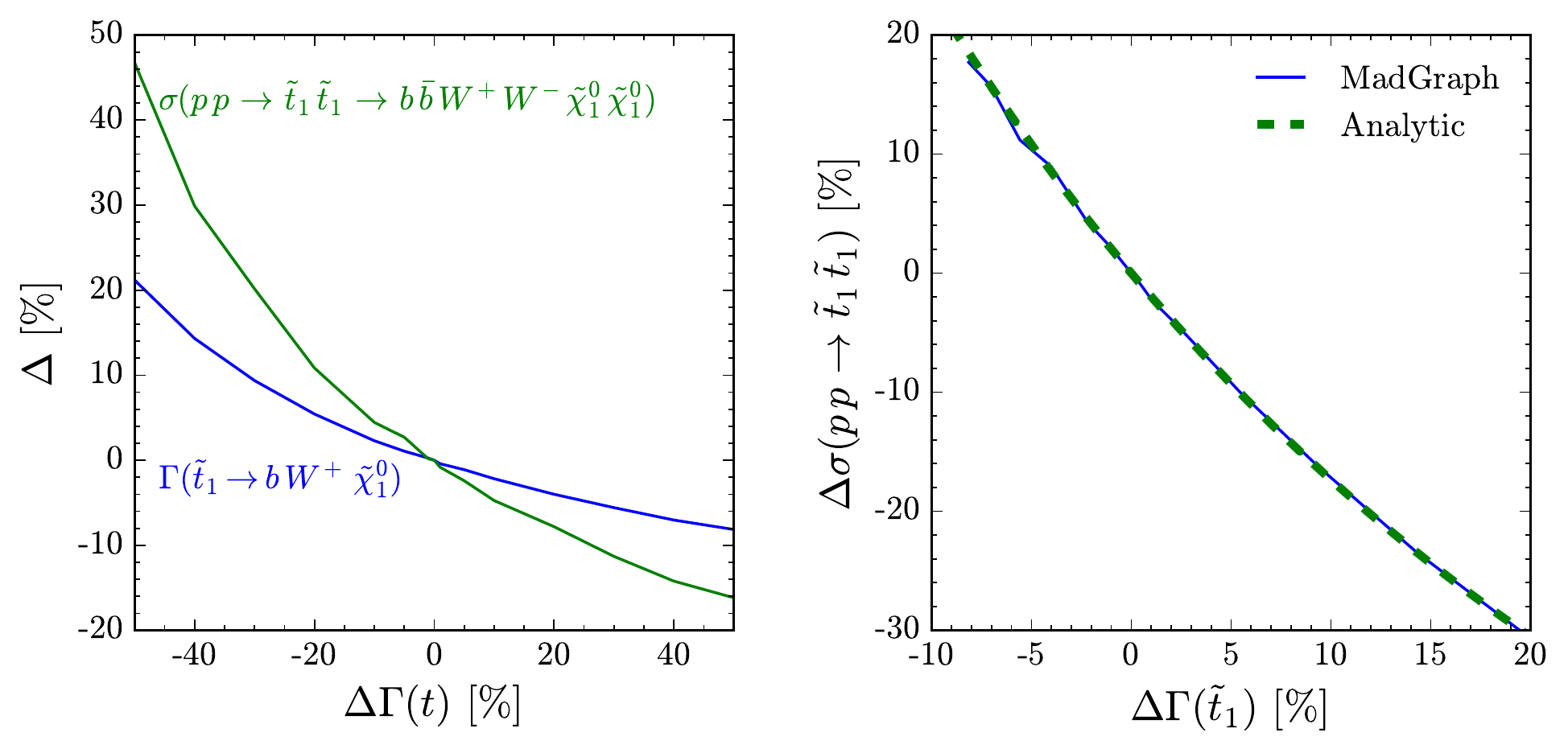}
\caption{The left panel shows the impact of the change in top width on the width of the stop (blue) and on the stop pair production cross section (green). The right panel shows  the impact of the change in the stop width on the stop pair production cross section (blue solid line); the corresponding analytic curve (green dashed line) is given in \cref{eqn:WidthErrorEstimate}.}
\label{fig:WrongWidths}
\end{center}
\end{figure}

This effect is illustrated numerically in \cref{fig:WrongWidths} using a stop mass near threshold. In the left panel, we adjust the the top width and the  blue line shows the extent to which the stop width changes as a result. The green line then shows the subsequent change in the production cross section.  The link between the stop width and cross section is made more obvious in the right panel; here we change the stop width directly and plot the resulting change in the production cross section. The blue line shows the numerical output from a \texttt{MadGraph} calculation while the green dashed line gives the analytic prediction demonstrating that \cref{eqn:WidthErrorEstimate} captures the dominant effect.

We see that in order to accurately estimate the cross section and efficiencies using the full matrix element, it is important to consistently calculate the widths. Due to the presence of two stop propagators in the production diagrams, inconsistencies in the widths lead to changes in the rates by $\sim2\, \Delta\Gamma / \Gamma$.

%**************** Section ***********************************
%\clearpage
\subsection{Spin correlations}
\label{sec:SpinCorr}
%*************************************************************

The last effect explored here, which is especially important for light stops near or below the top threshold, is the spin correlations of the decay products. When events are simulated in the NWA, \emph{i.e.}, as is usually done when decaying particles in \texttt{Pythia}, production is factorized from decay.  This implies that the angular distribution of the decay products do not include the correlations that result from the spin of the parent.  While it is possible to include matrix elements for decays in \texttt{Pythia 8}, the three-body decay of the stop is not currently implemented.  When $\texttt{Pythia}$ decays particles without a matrix element implementation, it assumes a constant matrix element and determines the kinematics strictly from phase space. 

\begin{figure}[t]
\begin{center}
\includegraphics[width=0.95\linewidth]{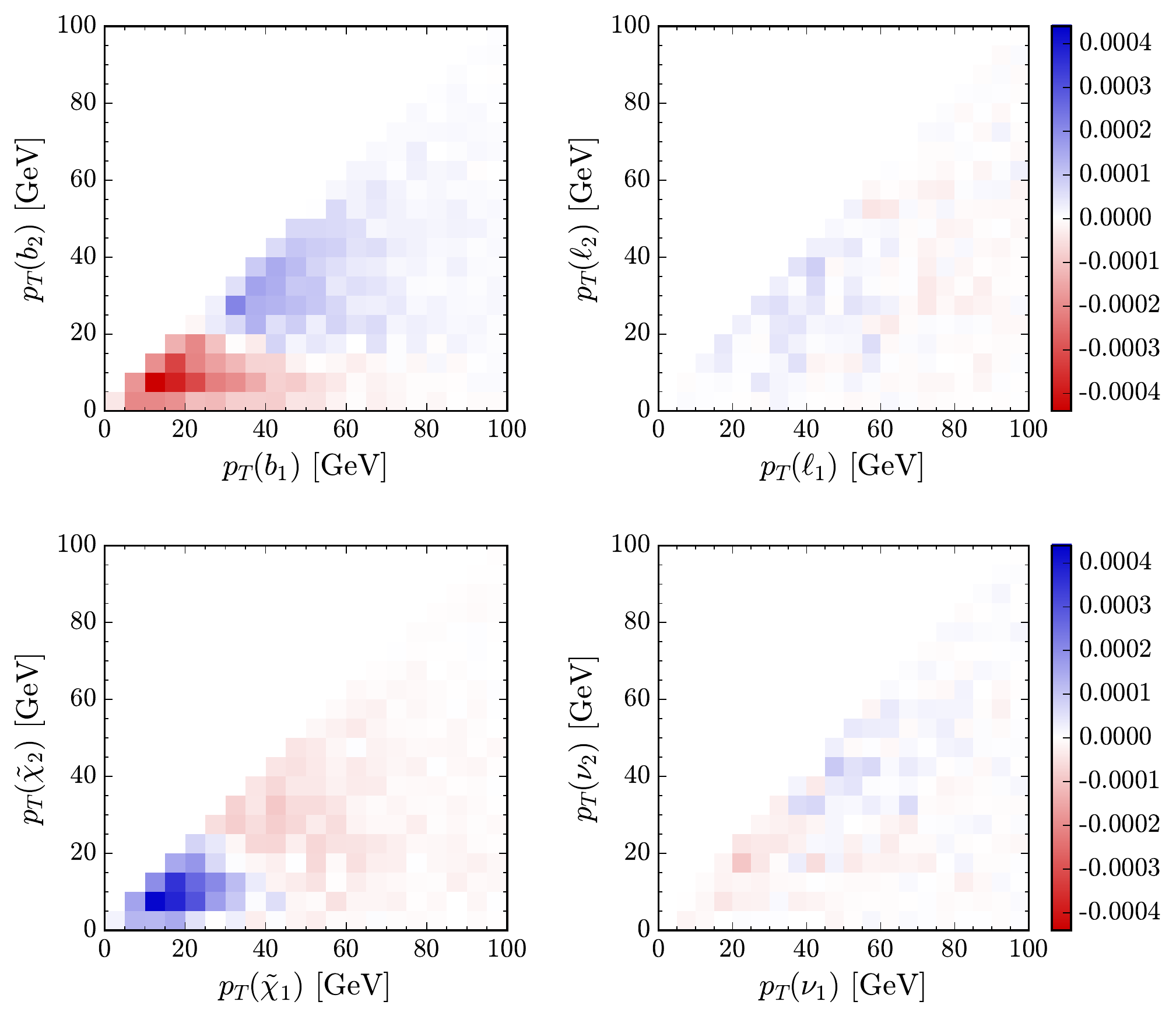}
\caption{Stop pair production events are generated with $\mstop = 150 \gev$ and $\mnino = 1\gev$ in the \mg\ and \pythia\ schemes (see \cref{tab:EvGenComparison} above for conventions).  The data sets are normalized and binned in the plane given by the $p_T$ of the second hardest particle versus the $p_T$ of the hardest particle for bottom quarks $b$, charged leptons $\ell$, the neutralino $\ninoone$, and the neutrino $\nu$.  The color scales show the result of subtracting the \pythia\ bins from the \mg\ bins. \pythia\ events do not include spin correlations which results in the $b$ quarks (upper left) being softer than they are when the full matrix element is used. This is compensated by harder neutralinos (lower left). There is some effect on the lepton and neutrino distributions, coming from the subsequent decay of the $W$, but the result is more subtle.}
\label{fig:Distributions}
\end{center}
\end{figure}

As an example for the importance of including spin correlations, \cref{fig:Distributions} examines the distributions of the final state particles for pair production of 150 GeV stops which decay to a 1 GeV neutralino, $b$ quark, lepton, and neutrino. We then compare the entire process, including decays, generated in \texttt{MadGraph} (\mg) with a calculation where the stops are decayed by \texttt{Pythia} (\pythia). The events are then binned according to the harder/softer truth-level $b$ quark, lepton, neutralino, and neutrino, normalized such that the sum of the bin content multiplied by the bin area is unity. The figure shows the result of taking the content of the \mg\ bins and subtracting the content of the \pythia\ bins. Bins with more \mg\ content are shaded blue, while the red bins contain more \pythia\ events.

Physically, the results of \cref{fig:Distributions} are due to the fact that the the outgoing particles for the \pythia\ events, are distributed according to phase space.  This neglects the fact that the $b$ and the $\ninoone$ are connected to the same scalar vertex and should therefore be correlated.  This results in $b$ quarks from the \pythia\ events being softer than from the \mg\ events. In contrast, the neutralinos are softer when the full matrix element is considered.

Now that we have reviewed a variety of physical effects that can impact the production calculation, we will move on to discuss how the ATLAS top cross section measurement is performed and how the limit on stealth stop production is derived.

%**************** Section ***********************************
%\clearpage
\section{Recast of the ATLAS Measurement and Search}
\label{sec:search}
%*************************************************************
This section begins with a review of the ATLAS measurement~\cite{TOPQ-2013-04} of the $t\bar{t}$ cross section that is used to probe the stealth stop region.  The non-trivial statistical procedure is explained, and a validation of our simulation framework is provided.  For details of the simulation framework, see \cref{sec:calculation} above.  

The ATLAS constraint on stealth stops~\cite{TOPQ-2013-04} is derived by placing a limit on the allowed contamination when measuring the $t\bar{t}$ cross section $\sigma_{t\bar{t}}$.  The approach is to extract $\sigma_{t\bar{t}}$ using both $\sqrt{s}=7 \tev$ and $8 \tev$ data in the $e^\pm\,\mu^\mp$ channel. An additional selection requiring either exactly one or two $b$-tagged jets is applied.  A simultaneous fit (which additionally allows the top mass to vary) utilizing all four bins is then performed to extract both the cross section and the $b$-tagging efficiency.  
The benefit of using a simultaneous fit is to minimize the effect of systematic uncertainties common to the cross-section measurements at the two masses. The result is $\sigma_{t\bar{t}}(8\tev) / \sigma_{t\bar{t}}(7\tev) = 1.326 \pm 0.057$, which is consistent with the Standard Model prediction~\cite{TOPQ-2013-04, Beneke:2011mq,Cacciari:2011hy,Czakon:2012zr,Czakon:2012pz,Czakon:2013goa,Czakon:2011xx}.

If one is willing to fix the top quark mass using independent observables, the measurements performed in~\cite{TOPQ-2013-04} can be interpreted as limiting the number of events coming from the pair production of stop squarks in the stealth stop region; see~\cite{Czakon:2014fka, Eifert:2014kea} for related pheno studies.  ATLAS uses this method to exclude a right-handed stop with masses between the top quark mass and 177 GeV, but for fixed neutralino mass of 1 GeV. We reinterpret their result to constrain regions of the $\mnino - \mstop$ plane, including stop masses below the top quark mass.\footnote{The ATLAS result was later extended in \cite{SUSY-2014-07} using the same analysis to including two parameter points with stop masses below the top mass.  Our approach differs in subtle ways as discussed below.}  We will present a statistical approach to handle the suite of correlated and un-correlated systematic errors, yielding a 95\% CL exclusion region using the CL$_s$ method.  As emphasized in the previous section, we also carefully include off-shell effects and spin correlations. 

%**************** Section ***********************************
\subsection{\boldmath Summary of the ATLAS $t\bar{t}$ Cross Section Measurement}
\label{sec:CrossSection}

This search relies on electrons, muons, and $b$-tagged jets with the following kinematic requirements.  An electron candidate must have a transverse momentum of $p_T(e^{\pm})> 25 \gev$ and pseudorapidity $ |\eta(e^{\pm})| < 2.47$. In addition, electrons near the transition region between the barrel and endcap $( 1.37 < |\eta(e^{\pm}) | < 1.52)$ are removed.  Muons are required to satisfy $p_T(\mu^{\pm}) > 25\gev$ and $|\eta(\mu^{\pm})| < 2.5$. Jet candidates must have $p_T(j) > 25 \gev$ and $|\eta(j)| < 2.5$.  We utilize the $b$-tagging efficiencies provided by \texttt{Delphes} for our stop signals; there is a different treatment of $b$-tagging for the $t\bar{t}$ prediction, as detailed in what follows. The events are triggered on either a single electron or muon, and the efficiencies of each of these triggers plateaus by a transverse momenta of 25 GeV.  

Each event considered in the measurement must have exactly one electron and one muon of opposite sign. Events are further classified as having exactly one or two $b$-jets (with no requirement on the number of non-$b$-tagged jets) resulting in $N_b$ and $N_{bb}$ number of events respectively.  The expected number of events in these channels is expressed as
\begin{equation}
\begin{aligned}
N_b & = \Lc ~ \sigma_{t\bar{t}}~ \epsilon_{e\mu} \, 2 \, \epsilon_b ~(1-C_b \, \epsilon_b) + N_b^{\text{bkg}} \\[5pt]
N_{bb} & = \Lc ~ \sigma_{t\bar{t}}~ \epsilon_{e\mu} \, C_b \, \epsilon_b^2 + N_{bb}^{\text{bkg}}\,,
\label{eqn:N1N2}
\end{aligned}
\end{equation}
where $\Lc$ is the integrated luminosity, $ \epsilon_{e\mu}$ is the efficiency for a $t\bar{t}$ event to pass the $e^\pm\,\mu^\mp$ preselection (including geometry and detector effects); ATLAS provides $\epsilon_{e\mu} = 7.7 \times 10^{-3}$. Additionally, $\epsilon_b$ is the efficiency to tag a $b$-jet, specifically in events coming from top quark decays. Then naively the probability of tagging the $b$-quarks coming from each of the two tops in an event would be $\epsilon_b^2$. However, kinematics and detector responses can lead to correlations for the two $b$-jets. This is taken into account by defining $C_b \equiv \epsilon_{bb} / \epsilon_{b}^2$, where $\epsilon_{bb}$ is the probability of tagging both $b$-quarks in a $t\bar{t}$ event. Finally, there are additional Standard Model processes that contribute to both $N_b$ and $N_{bb}$; these are denoted by $N_{b,bb}^{\text{bkg}}$.  This is dominated by single top production, which mainly contributes to $N_b^{\text{bkg}}$. 

The expressions for $N_{b,bb}$ in \cref{eqn:N1N2} can be solved for the $t\bar{t}$ cross section and the $b$-tagging efficiency, 
\begin{equation}
\begin{aligned}
\sigma_{t\bar{t}} &= \frac{C_b \, \Big(N_b - N_b^{\text{bkg}} + 2 \big(N_{bb} -N_{bb}^{\text{bkg}}\big) \Big)^2}{4 \, \Lc \, \big(N_{bb} - N_{bb}^{\text{bkg}} \big) \epsilon_{e\mu}} \\[10pt]
\epsilon_b & = \frac{2 \, \big( N_{bb} - N_{bb}^{\text{bkg}} \big)}{C_b \, \Big( N_b - N_b^{\text{bkg}} + 2\,\big(N_{bb} - N_{bb}^{\text{bkg}} \big) \Big)}\,.
\end{aligned}
\end{equation}
To extract $\sigma_{t\bar{t}}$ and $\epsilon_b$, the values of $C_b$ and $\epsilon_{e\mu}$ are estimated from simulations and given in \cref{tab:nuisance}. The cross section measurement is affected to a greater extent on the uncertainties from $N_b^{\text{bkg}}$ than from $N_{bb}^{\text{bkg}}$. Conversely, the uncertainty in $N_{bb}^{\text{bkg}}$ has a bigger impact on $\epsilon_b$. The measurement yields $\sigma_{t\bar{t}} = 182.9 \pm 7.1~\text{pb}$ at $\sqrt{s}=7\tev$ and $\sigma_{t\bar{t}} = 242.4 \pm 10.3~\text{pb}$ at $\sqrt{s}=8\tev$~\cite{TOPQ-2013-04}. The result of the two measured cross sections can then be used to infer the mass of the top quark; ATLAS gives the best fit as $m_t = 172.9^{+2.5}_{-2.6} \gev$~\cite{TOPQ-2013-04}.

%**************** Section ***********************************
\subsection{Validation of Constraint on Stealth Stop Production}

The $t\bar{t}$ cross-section measurement is sensitive to new physics that decays to top quarks.  As opposed to the cross section measurement, where $m_t$ is left as a free parameter, taking $m_t$ from other measurements allows the use of the predicted value of $\sigma_{t\bar{t}}$ to constrain the possibility of contamination from new physics. Using a top mass of $172.5 \pm 1.0$ GeV leads to top production cross sections of $177.3^{+11.5}_{-12.0}$ pb at $\sqrt{s}=7\tev$ and $252.9^{+15.3}_{-16.3}$ pb at $\sqrt{s}=8\tev$~\cite{TOPQ-2013-04,Beneke:2011mq,Cacciari:2011hy,Czakon:2012zr,Czakon:2012pz,Czakon:2013goa,Czakon:2011xx}. 

The ATLAS limits for stop pair production are set by simultaneously fitting the $7$ and $8\tev$ datasets and using profile likelihood ratios in the asymptotic limit~\cite{Cowan:2010js}. Correlated uncertainties are accounted for through the use of nuisance parameters. For instance, the dominant top quark pair production cross section uncertainties have a common origin, \emph{e.g.} the top mass uncertainty, and are thus treated as fully correlated. We take a simplified approach by modeling these uncertainties as Gaussians with widths of the averages of the upper and lower uncertainties for each distribution. To enforce that the 7 and 8 TeV predictions shift together, we introduce the nuisance parameter $\delta_{t\bar{t}}$ as:
\begin{align}
\sigma_{t\bar{t}}^7 &= \big(177.3  + 11.75 \times \delta_{t\bar{t}} \big) \pb\,; \notag \\[10pt] 
\sigma_{t\bar{t}}^8 &= \big(252.9  + 15.8 \times \delta_{t\bar{t}} \big) \pb\,.
\label{eq:sigmattbar}
\end{align}
Given the cross section, the number of expected events in a given bin coming from top quark pair production can be computed using the first terms in \cref{eqn:N1N2}. The additional nuisance parameters are given in \cref{tab:nuisance}.

In a similar manner, we treat the non-$t\bar{t}$ Standard Model contributions to the one and two $b$-tagged bins as correlated between the two energies.  ATLAS provides the number of expected events and size of the uncertainties in each of the bins. We translate this to cross sections, such that the uncertainty in the number of events is the result of the cross section and luminosity uncertainties added in quadrature. The resulting cross sections are then given by 
\begin{align}
\sigma_{N_b, \text{bkg}}^7 &= \big(86.96 + 8.55 \times \delta_{N_b} \big) ~\mathrm{fb}\,; \notag \\[10pt]
\sigma_{N_b, \text{bkg}}^8 &= \big(127.59 + 10.75 \times \delta_{N_b} \big) ~\mathrm{fb}\,; \notag \\[10pt]
\sigma_{N_{bb}, \text{bkg}}^7 &= \big(15.22 + 3.47 \times \delta_{N_{bb}} \big) ~\mathrm{fb}\,; \notag \\[10pt]
\sigma_{N_{bb}, \text{bkg}}^8 &= \big(22.66 + 6.37 \times \delta_{N_{bb}} \big) ~\mathrm{fb}\,.
\end{align}
where $\delta_{N_b}$ and $\delta_{N_{bb}}$ are the nuisance parameters.

Finally, we can compute the contribution from stop pair production to the signal regions. The central value of the stop production cross section is taken from \texttt{NLLFast}~\cite{Beenakker:2015rna}. The efficiencies of the selection criteria described in the previous section are estimated with the simulated events and are denoted by $\xi_{b}$ or $\xi_{bb}$ for the one and two $b$-tagged bins, respectively. The number of expected events from stop pair production in each of the four bins can then be given by
\begin{equation}
\big(N_{ \stopone \stopone}\big)_{(1,2)}^{(7,8)} = \Lc^{(7,8)} \, \sigma^{(7,8)}_{\stopone\stopone} \, \xi_{ (b,bb)}^{(7,8)}\,,
\end{equation}
where the superscripts denote the center of mass energy in TeV and the subscripts denote the number of $b$-tags.  The 7 and 8 TeV predictions are correlated through the cross section and luminosity, but each efficiency is treated as independent. As shown in~\cref{tab:nuisance}, we include a 10\% uncertainty on the stop production cross section. We have tested varying this between 5-20\% and find minimal changes in the final exclusions.

\begin{table}[t]
\centering
\renewcommand{\arraystretch}{1.7}
\setlength{\tabcolsep}{5pt}
\setlength{\arrayrulewidth}{3pt}
\begin{tabular}{l !{\vrule width 1.5pt} c c c c c c c c c c}
\hline
Parameter & $C_b^{7}$ & $C_b^{8}$ & $\epsilon_b^7$ & $\epsilon_b^8$ & $\Lc^7$& $\Lc^8$ & $\delta_{t\bar{t}}$ & $\delta_{N_b}$ & $\delta_{N_{bb}}$ & $\delta_{\tilde{t}_1 \tilde{t}_1}$ \\
Mean &  1.009 &  1.007 &  0.550 &  0.543 &  4.6 & 20.3 &  0 &  0 &  0 &  0  \\
Uncertainty &  0.0072 &  0.0063 &  0.0086 &  0.0058 &  0.0828 & 0.568 &  1 &   1 &   1 & 0.1 \\
\hline
\end{tabular}
\caption{Nuisance parameters used in our recast. The values of $C_b^{7,8}$, $\epsilon_b^{7,8}$, and $\Lc^{7,8}$ and their uncertainties are taken from the ATLAS analysis \cite{TOPQ-2013-04}. Each parameter is chosen from a Gaussian centered at the mean with a width given by the uncertainty.  The luminosity $\mathcal{L}$ has units of fb$^{-1}$ and all other parameters are dimensionless.}
\label{tab:nuisance}
\end{table}

Our framework uses likelihood ratios $L(\mu,\theta)$ to set limits on stop production, where $\mu$ is the signal strength parameter which is ultimately what gets constrained ($\mu=0$ is defined as Standard Model and $\mu=1$ is Standard Model + full stop signal), and $\theta$ are the nuisance parameters. The likelihood is defined as the product of the Poisson probabilities of the observed number of events in each of the signal regions $N_i^\text{obs}$ multiplied by the product of the Gaussian probabilities for yielding the specific value of the nuisance parameters, 
\begin{equation}
L\big(\mu, \mathbf{\theta} \big) = \Bigg(\prod_{i=1}^4 P\Big(N_i^{\text{obs}} \,\Big|\, \mu~N_{\tilde{t}_1\tilde{t}_1} + N_{t\bar{t}} + N_\text{bkg} \Big) \Bigg) ~ \Bigg( \prod_{j=1}^{n_{\text{nuis}}} G \Big( \theta_j \,\Big|\, \text{mean}_j, \text{var}_j \Big) \Bigg)\,,
\end{equation}
where $N_{\stopone \stopone}$ is the predicted number of stop pair production events, $N_{t\bar{t}}$ is the predicted number of top pair production events, $N_\text{bkg}$ is the predicted number of non-$t\bar{t}$ background events, and (mean$_j$, var$_j$) are given in \cref{tab:nuisance}.
The number of events observed by ATLAS are
\begin{equation}
N_b^7 = 3527, ~~N_{bb}^7 = 2073, ~~ N_b^8 = 21666, ~~\text{and}~~ N_{bb}^8 = 11739.
\end{equation} 

The likelihood is converted into a test statistic by taking the ratio of the constrained and unconstrained maximum likelihoods. In the unconstrained maximum, both the signal strength and the nuisance parameters are free to vary in order to maximize the likelihood. The values of the signal strength and nuisance parameters at this maximum are denoted by $\hat{\mu}$ and $\hat{\theta}$, respectively. The constrained maximum likelihood fixes the signal strength to a chosen value, and then the nuisance parameters roam to maximize the likelihood for that value of $\mu$; the values of the nuisance parameters in this case are given by $\hat{\hat{\theta}}$. Putting this together allows us to define
\begin{equation}
q_{\mu} = -2 \log  \frac{L\bigg(\mu, \hat{\hat{\theta}}\,\bigg)}{L\Big(\hat{\mu}, \hat{\theta} \Big)} .
\end{equation}
Large value of $q$ represent greater incompatibility with the data.  Next, we define a similar quantity under the assumption that only the Standard Model contributes to the data
\begin{equation}
q_{\mu,A} = -2 \log \frac{L\bigg(\mu, \hat{\hat{\theta}}\, \bigg)}{L\Big(0,\theta^{\prime} \Big)},
\end{equation}
where $\theta^{\prime}$ takes the central values for all of the nuisance parameters. 

Finally, the CL$_s$ method combines $q_\mu$ and $q_{\mu,A}$.  In the asymptotic limit of large number of events, the 95\% CL limit is set by solving for the value of $\mu$ which yields
\begin{equation}
\text{CL}_s(\mu) \equiv 0.05 = \frac{1 - \Phi\big(\sqrt{q_{\mu}}\big) }{\Phi \big( \sqrt{q_{\mu,A}} - \sqrt{q_{\mu}} \big)}
\label{eqn:cls95}
\end{equation}
where $\Phi$ is the cumulative distribution of the standard Gaussian with unit width.  For more information on this statistical procedure, see App.~A of~\cite{ATLAS:2011tau}.

%**************** Section ***********************************
%\subsection{Validation}
%*************************************************************

\begin{figure}[t]
\begin{center}
\includegraphics[width=0.5\linewidth]{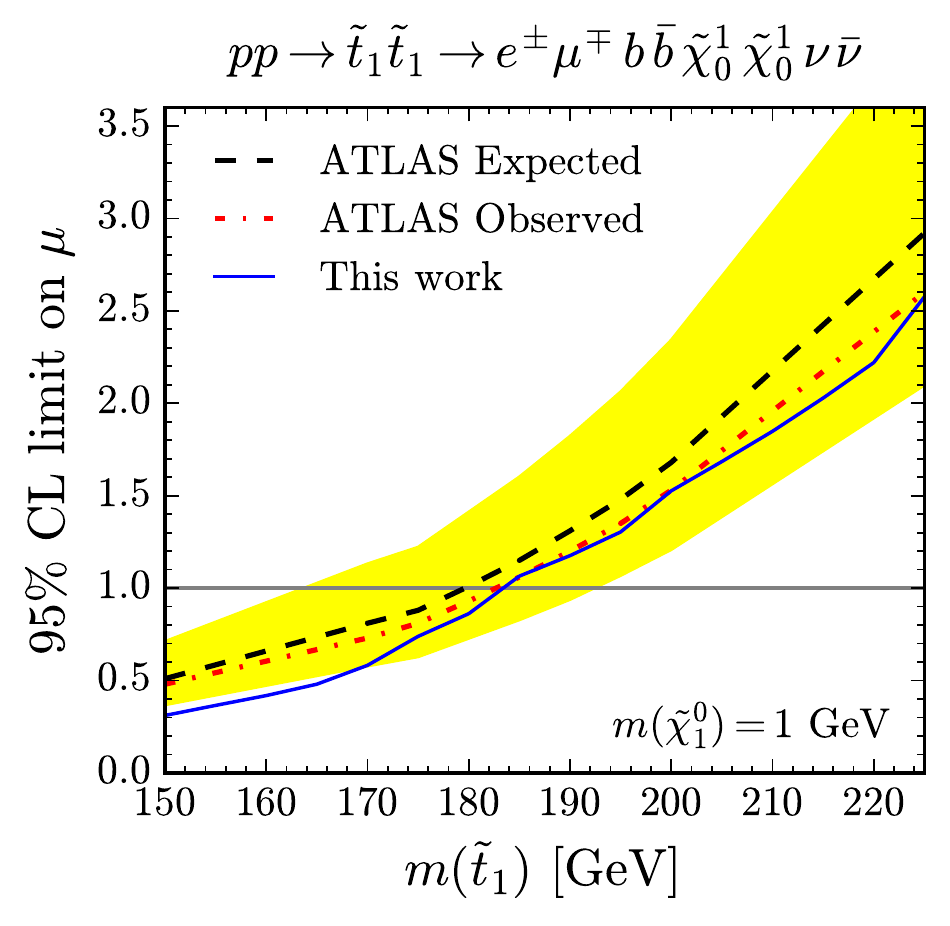}
\caption{Our recast of the ATLAS result~\cite{TOPQ-2013-04,SUSY-2014-07} of the signal strength $\mu$ vs. the stop mass; our result matches the observed limit over most of the parameter space. As the stop mass drops below 180 GeV, three-body effects and spin correlations start to impact the $b$-tagging efficiency, and our limit drops to the lower range of the expected uncertainty band as discussed in the text.}
\label{fig:ExclusionValidation}
\end{center}
\end{figure}

ATLAS placed limits on the stops assuming a neutralino mass of 1 GeV and a top mass of 172.5 GeV~\cite{TOPQ-2013-04,SUSY-2014-07}. In order to trust our extension of the analysis, we must first ensure that our framework can reproduce the ATLAS results to a good approximation. To this end, Fig.~\ref{fig:ExclusionValidation} shows the limit obtained by our analysis, \emph{i.e.}, the value of $\mu$ which satisfies \cref{eqn:cls95}, in comparison with the ATLAS results as a function of the stop mass.  The yellow uncertainty band is taken from ATLAS.  Values of $\mu = 1.0$ or less imply the stop pair production rate is excluded for that value of the mass. Over most of the region, our analysis follows very closely with the observed limit. 

We do note that our results start to deviate from the ATLAS observed limit (staying within the expected uncertainty band) below a stop mass of around 180~GeV.  This is not surprising as our statistical methods are different than those used by ATLAS for the signal.  ATLAS assumes that the $b$-tagging efficiency, $\epsilon_b$, is the same for both $t\bar{t}$ and $\stopone \,\stopone$ events. However, as noted in \cite{SUSY-2014-07}, for low stop masses, the ``fitted $\epsilon_b$ is different from what is expected for $t\bar{t}$ events alone''. Trying to keep the $t\bar{t}$ and the $\stopone\,\stopone$ efficiencies correlated in this regime (which is not done in our approach) is likely the reason for the difference. The best fit value of $\epsilon_b$ in our scenario (for $t\bar{t}$ alone) is shown in \cref{fig:NumEvents}.

\subsection{Impact of Spin Correlations and Finite Widths}

\begin{figure}[t]
\begin{center}
\includegraphics[width=0.95\linewidth]{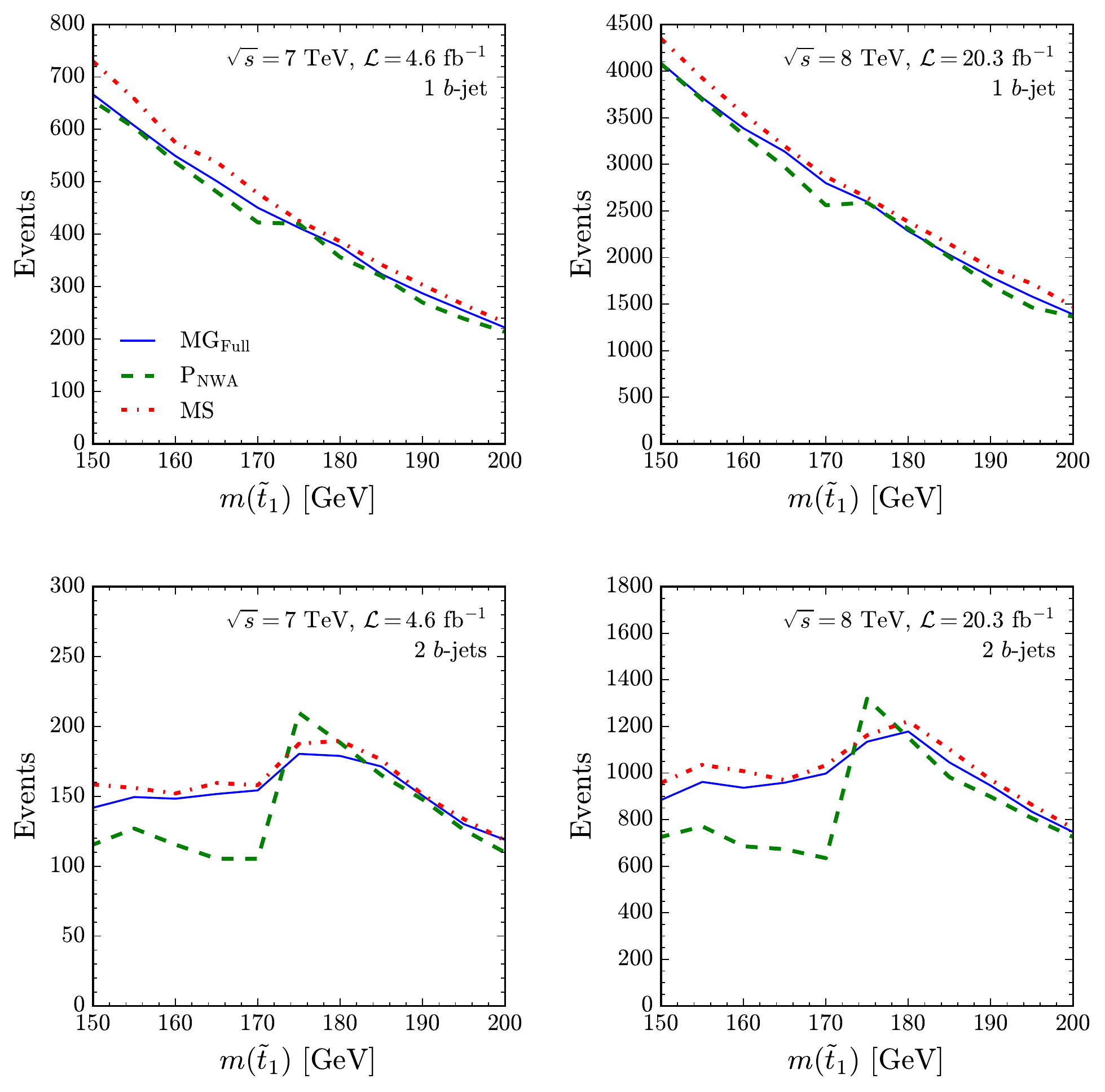}
\caption{The upper and lower panels show the number of events coming from stop pair production containing opposite sign $e\,\mu$ in the one and two $b$-tagged bins, respectively. The left and right panels show the two different center-of-mass energies, respectively.  The three event simulation approaches are summarized in \cref{tab:EvGenComparison}. }
\label{fig:EMUB}
\end{center}
\end{figure}

The impact of non-trivial spin correlations are investigated by examining each of the different event generation methods presented in \cref{tab:EvGenComparison}. The number of expected events coming from stop pair production in each of the four signal regions is shown in \cref{fig:EMUB} as a function of the stop mass. We sample the stop masses between 150-200 GeV with step sizes of 5 GeV. The blue, green-dashed, and red-dot-dashed lines represent whether the generation is preformed in the \mg, \pythia, and \ms\ approximations respectively. 

The upper panels show the one $b$-tag regions. In these, the number of expected events grows as the stop mass is decreased, due to the increase in the production cross section. The \mg\ and \ms\ predictions, which both include spin correlations, are not particularly sensitive to the position of the threshold as expected from \cref{fig:DecayWidths}. However, the \pythia\ line shows a dip at the threshold as it does not account for the three-body to two-body transition correctly. The two $b$-tag regions are displayed in the lower panel. The effect of the threshold is more dramatic, and the number of expected events is nearly constant below threshold for each method of decay. However, the \pythia\ approach yields $\sim 20\%$ less events. This is due to the lack of spin correlations for three-body stop decays in \texttt{Pythia}, which results in softer $p_T$ on average for the second $b$-jet.

\begin{figure}[t!]
\begin{center}
\includegraphics[width=0.5\linewidth]{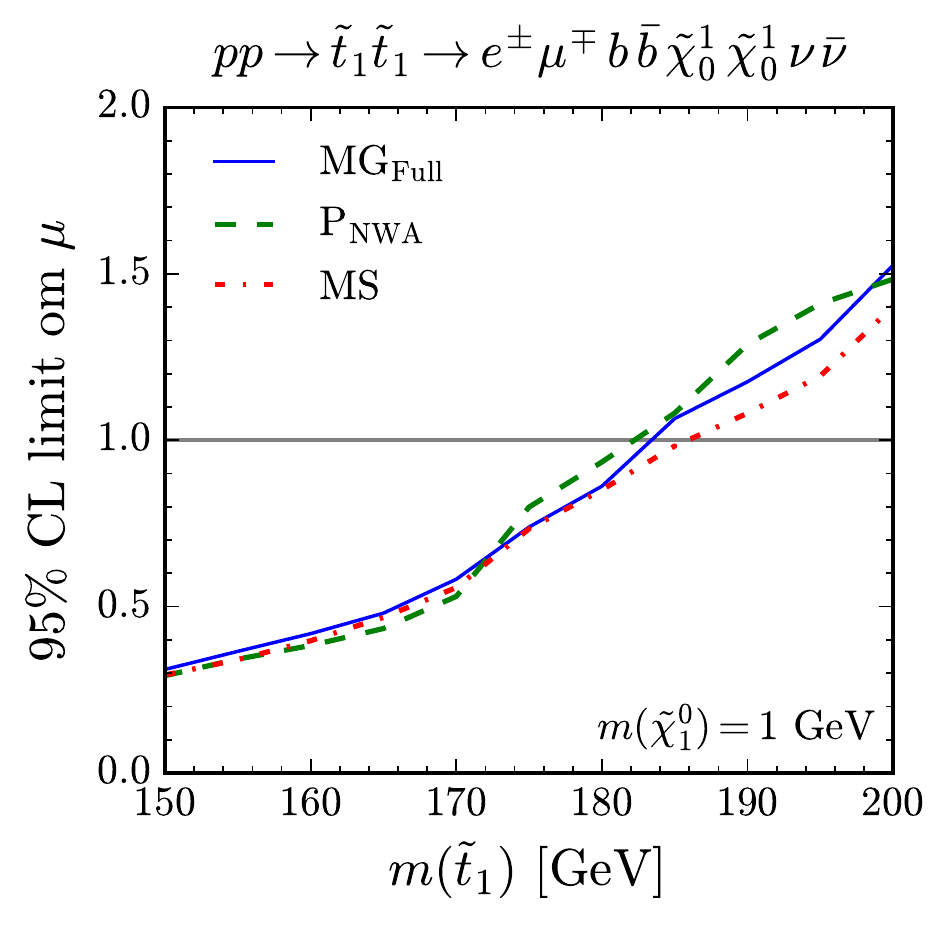}
\caption{We show the $95\%$ CL limit on the signal strength when the stops are decayed using the three approximations given in \cref{tab:EvGenComparison}.  Beyond the discontinuity around the threshold, there is little difference in the excluded values.}
\label{fig:Compare}
\end{center}
\end{figure}

The derived limits from the three different generation method are shown in \cref{fig:Compare}. The \ms\ line is slightly lower across the entire region. This is because there are slightly more events in each panel of \cref{fig:EMUB} due to the overestimated production cross section coming from the NWA, see \cref{sec:NWA}. Otherwise, since this approach models spin correlations, the results are very similar to the \mg\ exclusion. 

\begin{figure}[t]
\begin{center}
\includegraphics[width=0.95\linewidth]{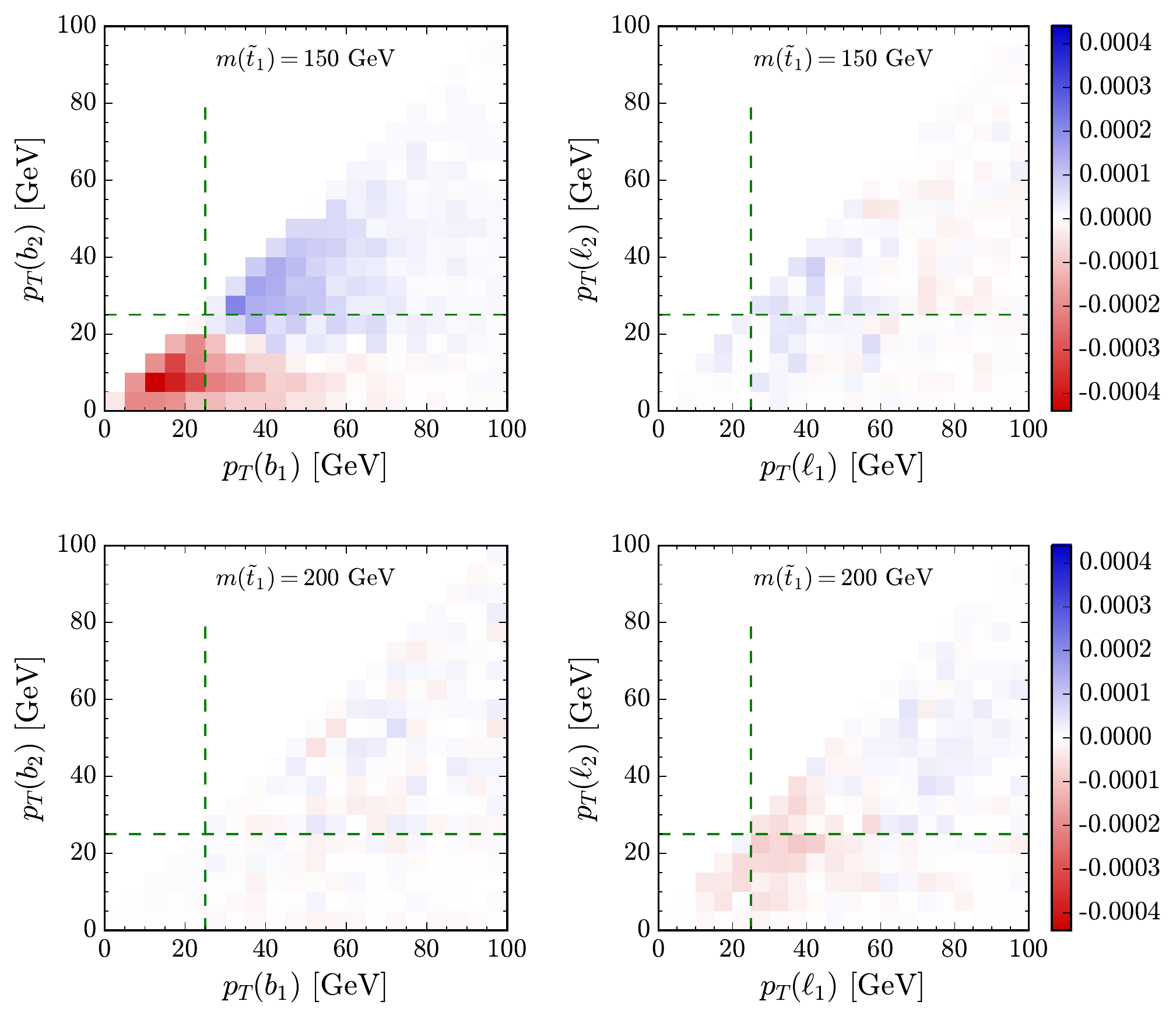}
\caption{Two dimensional histograms for \pythia - \mg \, for $m(\tilde{\chi}^0_1)=1\gev$; see the caption of \cref{fig:Distributions} for a detailed description.  The green dashed lines are at 25 GeV, the value of the cut used for event selection.  The mass of the stop in the upper panel is below the two-body threshold, while it is above threshold in the lower panel.}
\label{fig:SCSR}
\end{center}
\end{figure}

The \pythia\ exclusion line is less excluded at large masses, and then flips at the threshold to being more excluded than the approaches that include spin correlations. There are two reasons for this behavior.  Even though the cross section is over estimated due to the NWA, the lower right panel of \cref{fig:SCSR} shows that the softer lepton is more often too soft in comparison to the full calculation. This accounts for there being less events in each signal region above threshold in \cref{fig:EMUB}, which results in the weaker exclusion. The strong exclusion for the \pythia\ events below threshold is unintuitive.  As the upper left panel of \cref{fig:SCSR} shows, below threshold the second $b$-jet tends to be softer, leading to fewer events, as seen in \cref{fig:EMUB}.  It would seem that with fewer events, the exclusion should be weaker than the \mg\ line. However, the larger shift in the ratio of the one-and-two $b$-tagged regions is harder to accommodate in the fit, resulting in the stronger exclusion.

While all of the generation methods result in limits near a stop mass of 180 GeV, \cref{fig:EMUB} shows that there are thousands of extra events expected from the stops for masses larger than 180 GeV. The uncertainties impacting these regions are explored more in \cref{sec:Systematics}. It is surprising that the excluded signal strengths are so similar below threshold, even when the number of events are so different.

%**************** Section ***********************************
\section{Recasted Stealth Stop Results}
\label{sec:recast}
%*************************************************************

The results of our recast extends the ATLAS analysis to allow for a variety of neutralino masses.  In addition, we interpret the results in terms of both right- and left- handed stop pair production.

To cover the stealth stop splinter region, events are generate in a grid of the stop and neutralino masses given by:
\begin{equation}
\begin{aligned}
\mstop &\in \Big\{150, 155, 160, 165, 170, 175, 180, 185, 190, 195, 200\Big\}\gev,  \\[10pt]
\mnino &\in \Big \{0, 1\times10^{-4}, 1\times10^{-3}, 1, 5, 10, 15, 20, 25, 30, 35, 40, 45, 50, 55, 60\Big\} \gev \, . \nn
\end{aligned}
\end{equation}
For each parameter point, we generate 250,000 events at each center of mass energy for the full matrix element, and 1,000,000 events when using \texttt{Pythia} for the decays (because we do not specify the decay mode in this case). This is done both for right- and left-handed stops, which decay through $t_L$ and $t_R$, respectively.  Note that for the left-handed stops, we do not include a light left-handed sbottom in the spectrum.

\begin{figure}[htbp]
\begin{center}
\includegraphics[width=0.95\linewidth]{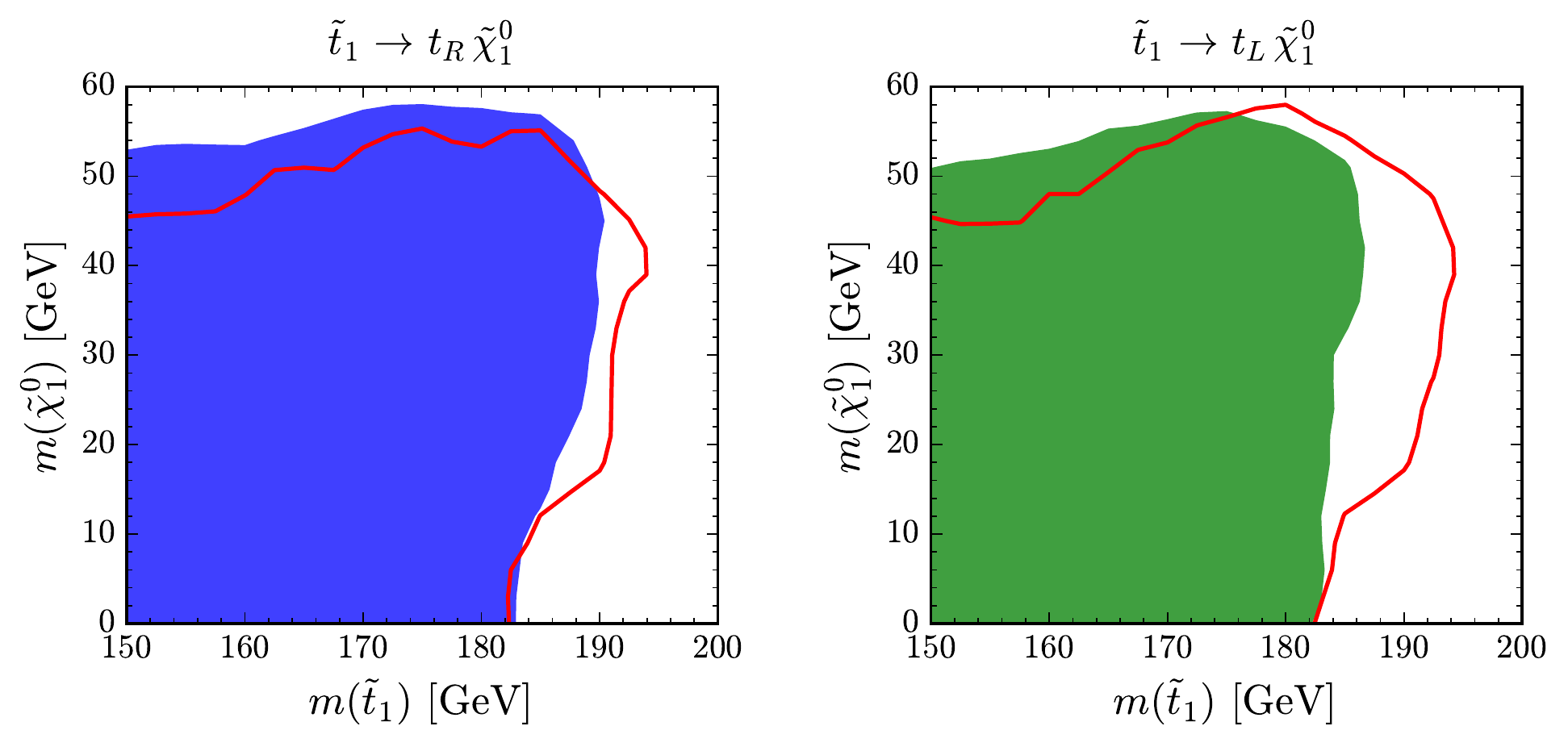}
\caption{The shaded regions are excluded by our reinterpretation of the ATLAS $t\bar{t}$  cross section measurement at the 95\% CL in the \mg\ approximation. The left (right) panel shows when the stop decays through a right- (left-) handed top. The red line shows what exclusion we would derive if spin correlations were not taken into account, \emph{i.e.} in the \pythia\ approximation.}
\label{fig:LeftRight}
\end{center}
\end{figure}

At each point in parameter space, the value of the signal strength which is excluded at $95\%$ CL$_s$ is computed, and a linear interpolation is used to extrapolate between the parameter points.  The shaded regions in \cref{fig:LeftRight} show which points are excluded for the full matrix element using the \mg\ approximation.  The left panel is for stops which decay through a right-handed top, and in the right panel, the stops decay through a left-handed top. The red line shows how the boundary shifts in the NWA without including the spin correlations, the \pythia\ approximation. 

These results show that neutralinos in the range $0 \gev< \mnino \lesssim 55\gev$ for $\mstop \lesssim 180\gev$ are excluded. There are only minor differences in the left- and right-handed stop exclusions when using the full matrix element with spin correlations. The limit extends smoothly between $\mnino = 1$ GeV, through the sub-GeV parameter space (which is a region of interest for new ideas in direct detection~\cite{Battaglieri:2017aum}), to $\mnino=0\gev$. Spin correlations were not critical for the validation plot \cref{fig:Compare} where $\mnino =  1 \gev$; the exclusions were very similar whether or not spin correlations were included. However, we see that in the full plane, spin correlations play a role. If they are not included, exclusions are too aggressive at large stop masses, pushing the would-be-excluded region to $\mstop \lesssim 190\gev$. In contrast, for lighter stops, the bounds are too conservative and do not cover the full area that is excluded when including more physics in the event generation.

%**************** Section ***********************************
%\clearpage
\subsection{Limits with Reduced Systematics}
\label{sec:Systematics}
%*************************************************************

\begin{figure}[t]
\begin{center}
\includegraphics[width=0.99\linewidth]{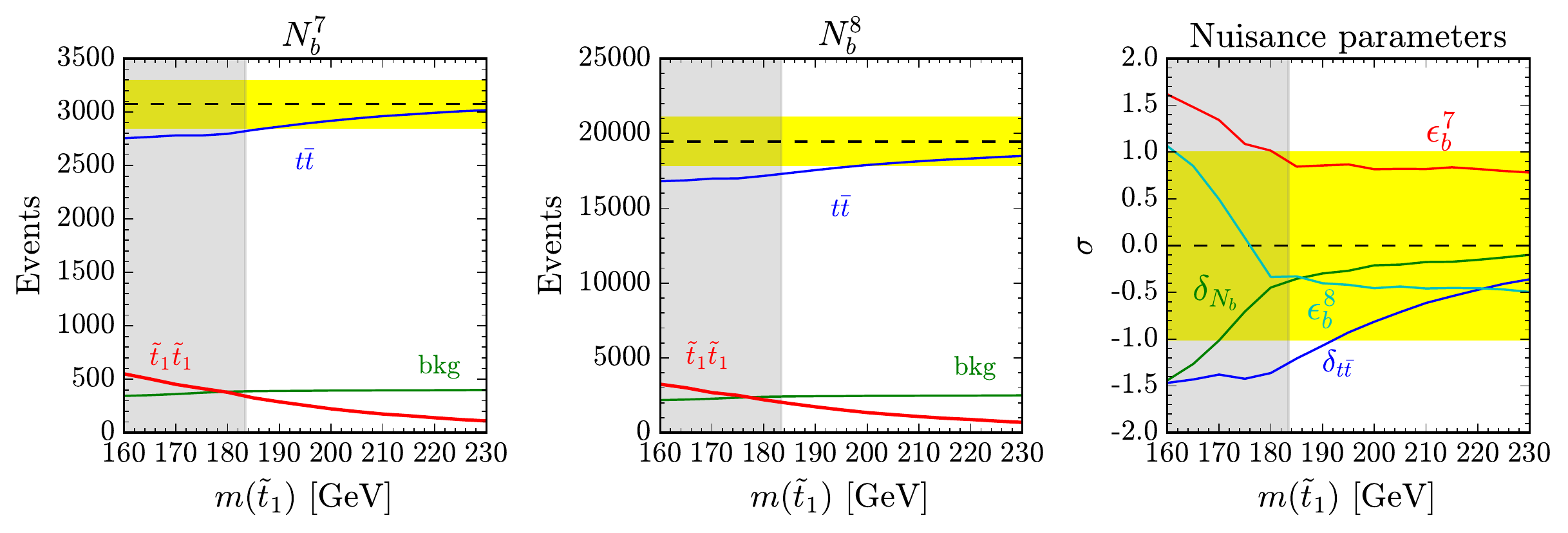}
\caption{The left and middle panels show the best fit number of expected events for a signal strength $\mu=1$ in the single $b$-tagged signal region for $\mnino =  1\gev$ at $\sqrt{s}=7\,,8\tev$ respectively. The blue, green, and red lines each show the contributions from $t\bar{t}$ production, other Standard Model backgrounds, and stop pair production, respectively. The dashed line shows the Standard Model expectation for the number from $t\bar{t}$ events and the yellow shaded region illustrates the one standard deviation uncertainty. The gray shaded regions are excluded. The right panel shows the number of standard deviations the nuisance parameters shift from their central values to maximize the likelihood.}
\label{fig:NumEvents}
\end{center}
\end{figure}

Our recast does not cover the entire stealth stop region; some of the splinter remains embedded.  In fact, it turns out that there can be thousands of events coming from stop production contributing to the signal region in the parameter space that we were not able to exclude. In this section, we examine the dominant systematic responsible for causing our limits to saturate and explore how the limits can improve if the uncertainty in the top cross section prediction were reduced.  In order to understand its quantitative impact, we have provided \cref{fig:NumEvents} which gives a sense of the relative size of the $\sigma_{t\bar{t}}$ uncertainty.
Rather than plotting the number of expected events, the \emph{best fit} values derived assuming a signal strength of $\mu = 1.0$ are shown. The left and middle plots display the $\sqrt{s}=7$ and 8 TeV one-$b$ tagged signal regions, respectively. The blue, green, and red lines denote the best fit values for the number of events coming from $t\bar{t}$ production, Standard Model backgrounds, and stop pair production. The yellow band shows the systematic error band for the number of $t\bar{t}$ events, which combines the top mass, pdf, and scale uncertainties. The black dashed line shows Standard Model expected value. The shaded region is excluded.

We see that in order to compensate for extra events coming from stops (as $\mstop$ becomes smaller), the $t\bar{t}$ best fit value can be driven below the error band. In particular, this is true for the best fit for the number of $t\bar{t}$ events in the gray excluded region. As the mass of the stop increases and the stop production cross section falls, the number of $t\bar{t}$ events trends toward the central value. The right panel helps expose what is driving these changes by showing the number of standard deviations away from the mean for the best fit values of a subset of the nuisance parameters. In order to accommodate the extra events from stops, the $t\bar{t}$ cross section is lowered, while  the $b$-tagging efficiencies are simultaneously being driven to larger values. At low stop masses, even the number of events coming from Standard Model backgrounds (which is a small contribution to the total number of events) are pushed outside their uncertainty band.

\begin{figure}[t]
\begin{center}
\includegraphics[width=0.95\linewidth]{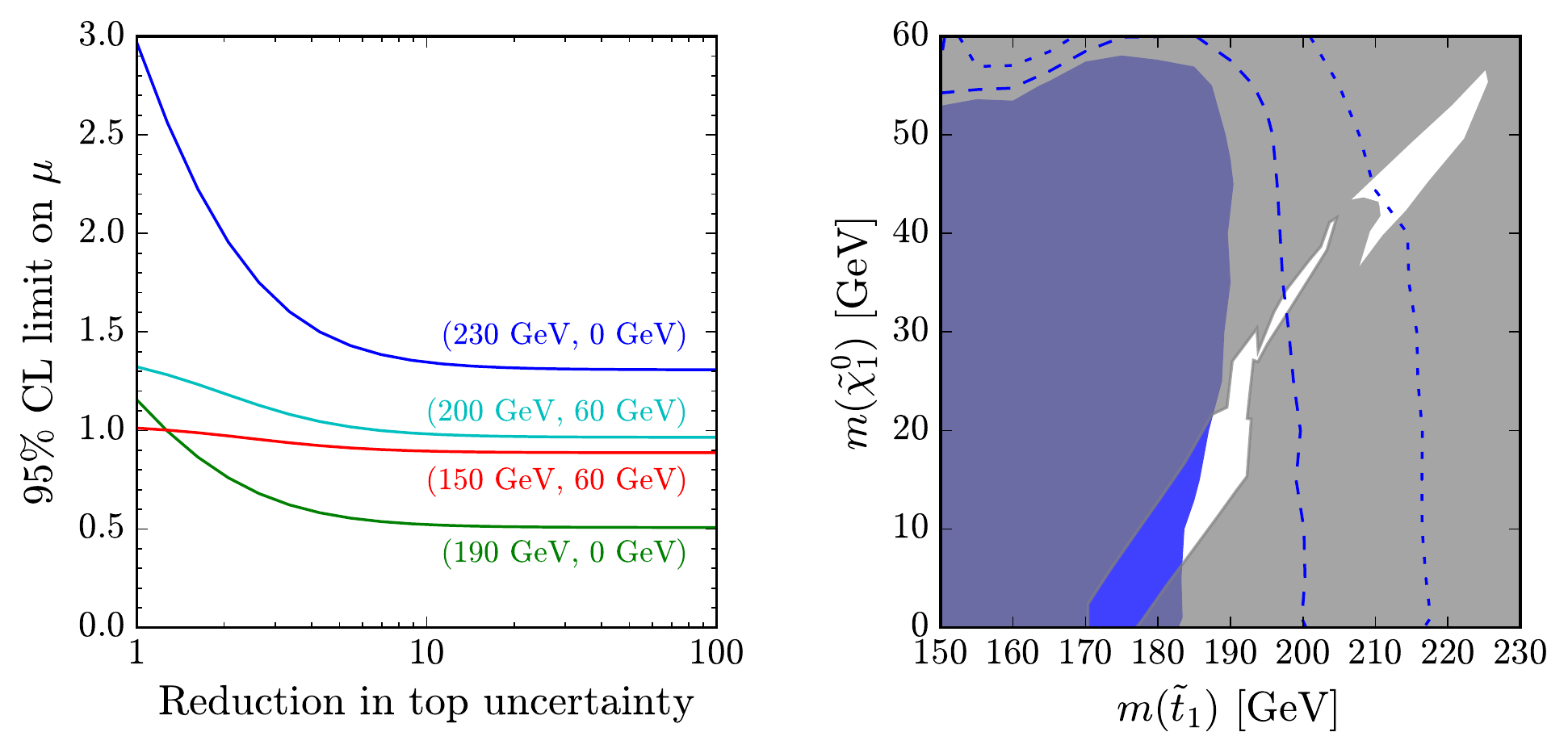}
\caption{The left panel shows the 95\% CL limit as a function of the $t\bar{t}$ uncertainty reduction for four points in the ($\mstop, \mnino$) parameter space.  The right panel shows how the limit in the $\mnino$ versus $\mstop$ plane could be improved if the uncertainty on the $t\bar{t}$ cross section were reduced by factors of 2 (dashed) and 10 (dotted).}
\label{fig:ExclusionSystematics}
\end{center}
\end{figure}

As shown in \cref{eq:sigmattbar}, the uncertainty on $\sigma_{t\bar{t}}$ is $\mathcal{O}(10\pb)$; the resulting uncertainty bands on the number of expected of events from $t\bar{t}$ production in \cref{fig:NumEvents} span many thousands of events. This explains why regions resulting in thousands of stop events are still not excluded.  Measuring the mass of the top quark to better accuracy using unrelated kinematic measurements, as well as improving the pdf and scale uncertainties, could reduce this systematic, thereby yielding a stronger limit. To illustrate the potential impact this could have, the left panel of \cref{fig:ExclusionSystematics} shows the value of the signal strength that could be excluded at the $95\%$ CL for a few different mass points in our parameter scan as a function of the improvement factor for the $t\bar{t}$ cross-section uncertainty. For the massless neutralino parameter points (blue and green), the value of the signal strength which is excluded drops very quickly as the $t\bar{t}$ cross-section uncertainty is reduced. The green line, with $\mstop = 190\gev$, goes from being allowed to excluded by reducing the uncertainty by less than a factor of 2. However, the exclusion on the signal strength for the $\mstop=230\gev$ parameter point never gets below 1, so it cannot be ruled out by reducing the $t\bar{t}$ cross-section uncertainty alone simply due to the lack of raw signal events.

The red and cyan lines are for a heavier neutralino $\mnino=60\gev$. The improvement on the excluded value of the signal strength does not change as dramatically when the $t\bar{t}$ cross-section uncertainty is reduced. This occurs because the $b$-jets are much softer  when the neutralinos are heavier, and as such do not pass the selection cuts as often.  Intuitively, the size of the $t\bar{t}$ cross-section uncertainty does not have a large impact when so few stop events pass the selection cuts.

The right panel of \cref{fig:ExclusionSystematics} shows the exclusion computed here overlaid on the stop splinter region. The blue shaded region is the same as in \cref{fig:LeftRight}. The dashed line shows how the excluded region would change if the uncertainty on the $t\bar{t}$ cross section were reduced by a factor of 2, while keeping the rest of the analysis fixed.  If this uncertainty could be reduced, a large portion of the lower sliver of space not yet excluded by ATLAS could be covered. The dot-dashed line shows the projected exclusion with a factor of 10 reduction of the $t\bar{t}$ cross-section uncertainty, demonstrating that it is in principle possible to cover the entire lower sliver.  While it is unlikely that this approach could ever be used to constrain the upper part of the open region, a reasonable expectation is that complimentary direct searches for stops will play an important role as more data is analyzed.

%**************** Section ***********************************
%\clearpage
\section{Discussion}
\label{sec:discussion}
%*************************************************************

In this work, we have examined the light stop splinter region not yet excluded by ATLAS.  We showed the impact of including both finite-width effects and spin correlations.  Our work shows that the splinter region can be excluded for $\mstop \lesssim 180 \gev$.  We also demonstrated that a reduction in  the uncertainty on the top quark production cross section by a factor of 2 would raise the limit on $\mstop$ by around 20 GeV. This would go a log way toward closing the lower splinter region, leaving the upper part to be closed by dedicated stop searches. In addition, ATLAS typically uses $\mnino=1\gev$ for the lightest point in their exclusion scans. We verified that the limits smoothly extend to $\mnino=0\gev$.

When performing this study, careful attention was needed due to some of the assumptions build into the event generation tools.  It is very common when calculating the decay width to assume that the three-body effects are not important if two-body modes are open.  While this is often a very good approximation, it is simply not true near the two-body to three-body threshold. Furthermore, the decay widths of the stops in this region are on the order of 0.1-10 MeV; \texttt{MadGraph} recognizes that these widths are smaller than the QCD scale and as a result treats the stops as stable by setting their width to zero.  The result is that the stop cannot decay and event generation fails.  It does do this for good reason since colored particles, such as the stop, with widths this small could form bound states like stoponium.  However, they also decay a large fraction of the time before this happens; see~\cite{Martin:2008sv,Martin:2009dj,Younkin:2009zn,Kumar:2014bca,Batell:2015zla} for more information and limits on stoponium.  A detailed workaround is presented in \cref{sec:eventgen}.

Everywhere in this paper, we assumed the LSP is a bino-like neutralino.  It would also be interesting to reinterpret this search assuming the LSP is a gravitino.  For much of the parameter space, the stops have small enough widths that they would live long enough to leave the detector.  However, if one imagines that the stop width were larger, \emph{e.g.} due to the presence of another decay mode, it would be interesting to study the impact of the spin correlations for this scenario. Finally, generalizing the exclusion to the $m_t$-$\mstop$-$\mnino$ parameter space (as was done in \cite{Czakon:2014fka}) is not possible within our framework since we do not simulate the top and Standard Model backgrounds.
We leave such explorations to future work.

%**************** Section ***********************************
\section*{Acknowledgments}

We are grateful to S. Martin for useful discussions about stoponium. 
TC is supported by the U.S. Department of Energy under grant number DE-SC0018191.  WH and SM are supported by the U.S. Department of Energy under grant number DE-SC0012008.  BO is supported by the U.S. Department of Energy under grant number DE-SC0011640.  This work utilized the University of Oregon Talapas high performance computing cluster.

%*************************************************************
\appendix
\section*{Appendices}
%*************************************************************

%**************** Section ***********************************
\section{Stealth Stop Event Generation with \texttt{Madgraph}}
\label{sec:eventgen}
%*************************************************************

Near threshold, care must be taken when generating events in order to include all finite width effects and spin correlations. We compute the matrix elements using \texttt{MadGraph}, without demanding that any particle (the top quark or $W$) appear on shell, which allows for the same process to be used above and below threshold. In a generic MSSM point, there are many diagrams which could contribute to such a process, however we explicitly forbid other sparticles to appear in the diagrams because we are interested in simplified models. We use the following process in order to only get leptonic events (with taus included in the definition).
\begin{lstlisting}[basicstyle=\ttfamily\small]
import model MSSM_SLHA2
define susy = t2 t2~ b1 b1~ b2 b2~ n2 n3 n4 x1+ x1- x2+ x2- h2 h3 h+ h-
define l = e+ e- mu+ mu- ta+ ta- vl vl~
generate p p > t1 t1~, (t1 > b n1 l l / susy), (t1~ > b~ n1 l l  / susy)
\end{lstlisting}
 As an added complication, \texttt{compute\_widths} actually fails for most of the parameter space in our scan because the width of the stop is small compared the the QCD scale, so it sets the width to 0.\footnote{Such a small width does imply there is the additional possibility of producing stoponium instead of stop pairs~\cite{Martin:2008sv,Martin:2009dj,Younkin:2009zn,Kumar:2014bca,Batell:2015zla}.} To get a reliable value for the stop width, we use the following process in \texttt{MadGraph}.
\begin{lstlisting}[basicstyle=\ttfamily\small]
import model MSSM_SLHA2
define susy = t2 t2~ b1 b1~ b2 b2~ n2 n3 n4 x1+ x1- x2+ x2- h2 h3 h+ h-
define f = u u~ d d~ s s~ c c~ b b~ e+ e- mu+ mu- ta+ ta- vl vl~
generate t1 > b n1 f f / susy
\end{lstlisting}
This decay process computes the matrix element including full propagators for the top and $W$, and therefore critically depends on the definition of both of their widths in the parameter card. It is important that the widths chosen for the top and $W$ are consistent between the two different \texttt{MadGraph} processes (the decay and production); we do this using \texttt{compute\_widths 6 24}, such that the top and $W$ width are consistently defined compared to the model parameters and not necessarily equal to the PDG values. 

%**************** Section ***********************************
\section{Stealth Stop Event Generation Pitfalls}
\label{sec:pitfalls}
%*************************************************************

In \cref{sec:calculation}, we detailed the need for using correct widths, and showed that common calculators actually get the width wrong near threshold. This forces the user to compute the width and production in completely independent processes. In order to get correct results, the parameters used between these two processes needs to be consistent. Needing to do these separately introduces the extra possibility of user error. This appendix details some of the pitfalls we ran into along the way and should serve as a guide to correct calculations.

\begin{figure}[t]
\begin{center}
\includegraphics[width=0.95\linewidth]{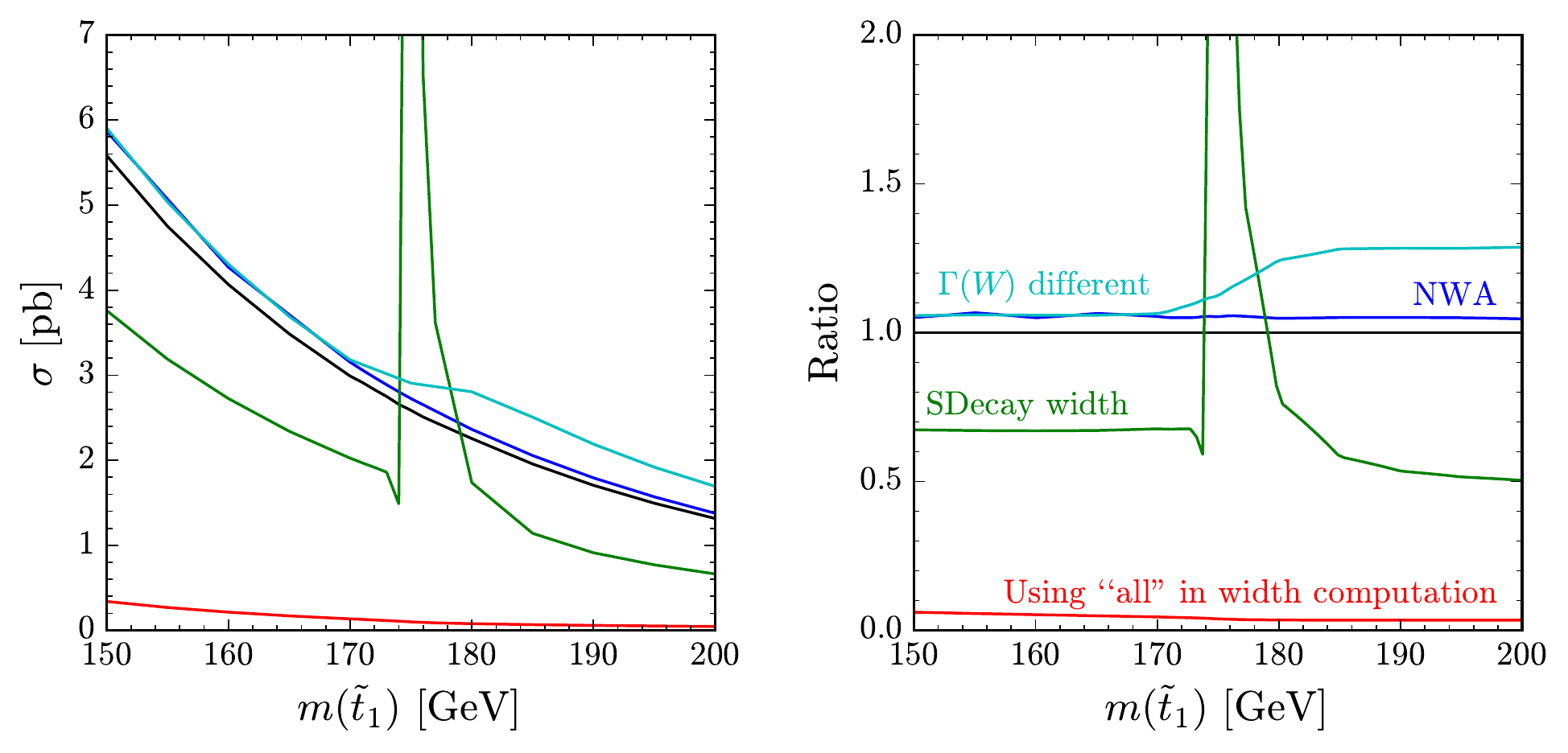}
\caption{The left panel shows the production cross section at $\sqrt{s}=8$ TeV for the process $p\,p\rightarrow \stopone \,\stopone \rightarrow b \,\bar{b}\, \ell^+ \,\ell^-\, \nu_{\ell}\, \bar{\nu}_{\ell}\, \ninoone\,\ninoone$, while the right panel shows the ratio compared to the method used in this paper. The black lines show the value when the stop width is computed to the final state particles. The blue lines are the narrow width approximation, wherein the total stop production cross section is multiplied by branching ratio for each $W$ to decay leptonically. The green lines correspond to using the width obtained from \texttt{SDecay}, which includes NLO QCD corrections. The red and light blue lines come from common user errors.}
\label{fig:BadWidth}
\end{center}
\end{figure}

In Fig.~\ref{fig:BadWidth}, the importance of using consistent widths is demonstrated. The left panel shows the cross section for stop pair production in the dileptonic channel at $\sqrt{s}=8$ TeV, and the right panel displays the ratio of different methods to our procedure (denoted in black). The results of using the width computed by \texttt{SDecay} are shown in green. As \texttt{SDecay} includes NLO QCD corrections, the width in the denominator of the propagators is larger than expected from the couplings, which leads to a production cross section which is too small. Just above threshold, \texttt{SDecay} only uses the 2-body channel, yielding too small of stop widths and extra large cross sections.

The need to account for more than just the two-body width, along with having to compute the width and production separately, opens the possibility of extra (user) inconsistencies. For instance, in generating the decay width of the stops through to the final state particles, it is tempting to use
\texttt{generate t1 > b n1 all all / susy},
expecting that the two ``all''s account for the $W$ decay. However, in addition to the $W$ decay products, this also produces diagrams with an un-decayed $W$ and radiated gluons. Because of these extra diagrams, the result of the process is not the width of the stop; the width is much too large. This leads to reduced cross sections, as shown in red in Fig.~\ref{fig:BadWidth}. Alternatively, the cyan lines show the effect of using different widths for the top and $W$ between calculating the stop decay and the production. For this, the \texttt{compute\_widths 6 24} command is used when getting the stop width, but the user ``forgets'' to update the top and $W$ width from the (slightly larger) default values in the production process. In order to get valid results, the matrix element for the stop width needs to be consistent with that used in the production and subsequent decay.

As a last consistency check, we compare our leptonic cross section with the full stop pair production cross section (\texttt{p p > t1 t1\~}). Using the narrow width approximation (NWA) we take the total cross section and multiply by the leptonic branching ratio (of the $W$s), and show this in the blue lines of Fig.~\ref{fig:BadWidth}. All of the collider level cuts in the \texttt{run\_card.dat} are removed for the full (production and decay) process to get the most accurate comparison. With this, the NWA is larger than our result by $\sim5-6\%$ across all stop masses, consistent with our observations in \cref{sec:NWA}. 

To get a further sense of what causes the differences, we show the values of the widths and corresponding cross section for the point $\mstop = 200\gev$ in \cref{tab:xsecExamples}. In the first block, the widths of the $W$, $t$, and $\stopone$ are given, underscored by the method for which they are calculated. The $t$ widths marked by three-body implicitly use the given $W$ width. Similarly, the $\stopone$ widths are calculated all the way to the final state leptons, and use the corresponding $W$ and $t$ widths. In doing so, the $\stopone$ width used in production process is consistent and therefore yields the appropriate cross section. This is exemplified in the row where the $W$ and $t$ widths are set to 1.0, and the stop is calculated consistent with that. The resulting cross section remains the same.  In the last block of numbers, the stop width is not consistent. 

\begin{table}[p]
\centering
\renewcommand{\arraystretch}{2}
\setlength{\tabcolsep}{15pt}
\setlength{\arrayrulewidth}{3pt}
\begin{tabular}{c c c!{\vrule width 1.5pt} c !{\vrule width 1.5pt} c}
\hline
$\Gamma(W)$	&	$\Gamma(t)$	&	$\Gamma({\footnotesize \stopone})$	&	$\sigma_{\text{leptonic}}$	&	Notes \\
\noalign{\hrule height 1.5pt}
N/A 	& 	N/A 	&	N/A  & 1.37 & Narrow width \\
\noalign{\hrule height 1.5pt}
$\underbrace{2.085}_{\text{PDG}}$ & $\underbrace{1.41}_{\text{PDG}}$ & 0.02248 & 1.312 & PDG values \\
$\underbrace{2.143}_{\text{2-body}}$ & $\underbrace{1.557}_{\text{2-body}}$ & \textcolor{blue}{0.01979} & 1.310 & \texttt{compute\_widths 6 24} \\
$\underbrace{2.085}_{\text{PDG}}$ & $\underbrace{1.575}_{\text{3-body}}$ & 0.02012  & 1.310 & only $W$ PDG \\
$\underbrace{2.143}_{\text{2-body}}$ & $\underbrace{1.532}_{\text{3-body}}$ & 0.02013  & 1.310 & Completely consistent\\ [15pt]
\noalign{\hrule height 1.5pt}
1.0 & 1.0 & 0.06674 & 1.312 & Wrong but consistent \\
\noalign{\hrule height 1.5pt}
$\underbrace{2.085}_{\text{PDG}}$ & $\underbrace{1.41}_{\text{PDG}}$ & $\underbrace{\textcolor{red}{0.01979}}_{\text{from \textcolor{blue}{blue}}}$ & 1.694 & $\Gamma(W)$ different\\
$\underbrace{2.143}_{\text{2-body}}$ & $\underbrace{1.557}_{\text{2-body}}$ & \textcolor{red}{0.1072} & 0.04 & Used ``all'' instead of ``f'' \\
2.003 & 1.384 & 0.03354 & 0.6627 & \texttt{SDecay} widths \\
\hline
\end{tabular}
\caption{Cross section in pb for different widths in GeV of $t$, $W$, and $\stopone$. We set $m_t = 173.1 \text{ GeV}$, $\mstop = 200$ GeV, and $\mnino = 1$ GeV.}
\label{tab:xsecExamples}
\end{table}

\clearpage

\bibliographystyle{utphys}
\bibliography{Degenerate}

\providecommand{\href}[2]{#2}\begingroup\raggedright\begin{thebibliography}{10}

\bibitem{Martin:1997ns}
S.~P. Martin, ``{A Supersymmetry primer},''
  \href{http://arxiv.org/abs/hep-ph/9709356}{{\tt arXiv:hep-ph/9709356
  [hep-ph]}}.
[Adv. Ser. Direct. High Energy Phys.18,1(1998)].
%%CITATION = HEP-PH/9709356;%%.

\bibitem{Giudice:2017pzm}
G.~F. Giudice, ``{The Dawn of the Post-Naturalness Era},''
\href{http://arxiv.org/abs/1710.07663}{{\tt arXiv:1710.07663
  [physics.hist-ph]}}.
%%CITATION = ARXIV:1710.07663;%%.

\bibitem{Dimopoulos:1995mi}
S.~Dimopoulos and G.~F. Giudice, ``{Naturalness constraints in supersymmetric
  theories with nonuniversal soft terms},''
  \href{http://dx.doi.org/10.1016/0370-2693(95)00961-J}{{\em Phys. Lett.} {\bf
  B357} (1995)  573--578},
\href{http://arxiv.org/abs/hep-ph/9507282}{{\tt arXiv:hep-ph/9507282
  [hep-ph]}}.
%%CITATION = HEP-PH/9507282;%%.

\bibitem{Cohen:1996vb}
A.~G. Cohen, D.~B. Kaplan, and A.~E. Nelson, ``{The More minimal supersymmetric
  standard model},''
  \href{http://dx.doi.org/10.1016/S0370-2693(96)01183-5}{{\em Phys. Lett.} {\bf
  B388} (1996)  588--598},
\href{http://arxiv.org/abs/hep-ph/9607394}{{\tt arXiv:hep-ph/9607394
  [hep-ph]}}.
%%CITATION = HEP-PH/9607394;%%.

\bibitem{Papucci:2011wy}
M.~Papucci, J.~T. Ruderman, and A.~Weiler, ``{Natural SUSY Endures},''
  \href{http://dx.doi.org/10.1007/JHEP09(2012)035}{{\em JHEP} {\bf 09} (2012)
  035},
\href{http://arxiv.org/abs/1110.6926}{{\tt arXiv:1110.6926 [hep-ph]}}.
%%CITATION = ARXIV:1110.6926;%%.

\bibitem{Brust:2011tb}
C.~Brust, A.~Katz, S.~Lawrence, and R.~Sundrum, ``{SUSY, the Third Generation
  and the LHC},'' \href{http://dx.doi.org/10.1007/JHEP03(2012)103}{{\em JHEP}
  {\bf 03} (2012)  103},
\href{http://arxiv.org/abs/1110.6670}{{\tt arXiv:1110.6670 [hep-ph]}}.
%%CITATION = ARXIV:1110.6670;%%.

\bibitem{Buckley:2016tbs}
M.~R. Buckley, A.~Monteux, and D.~Shih, ``{Precision Corrections to Fine Tuning
  in SUSY},'' \href{http://dx.doi.org/10.1007/JHEP06(2017)103}{{\em JHEP} {\bf
  06} (2017)  103},
\href{http://arxiv.org/abs/1611.05873}{{\tt arXiv:1611.05873 [hep-ph]}}.
%%CITATION = ARXIV:1611.05873;%%.

\bibitem{Evans:2013jna}
J.~A. Evans, Y.~Kats, D.~Shih, and M.~J. Strassler, ``{Toward Full LHC Coverage
  of Natural Supersymmetry},''
  \href{http://dx.doi.org/10.1007/JHEP07(2014)101}{{\em JHEP} {\bf 07} (2014)
  101},
\href{http://arxiv.org/abs/1310.5758}{{\tt arXiv:1310.5758 [hep-ph]}}.
%%CITATION = ARXIV:1310.5758;%%.

\bibitem{Papucci:2014rja}
M.~Papucci, K.~Sakurai, A.~Weiler, and L.~Zeune, ``{Fastlim: a fast LHC limit
  calculator},'' \href{http://dx.doi.org/10.1140/epjc/s10052-014-3163-1}{{\em
  Eur. Phys. J.} {\bf C74} (2014) no.~11, 3163},
\href{http://arxiv.org/abs/1402.0492}{{\tt arXiv:1402.0492 [hep-ph]}}.
%%CITATION = ARXIV:1402.0492;%%.

\bibitem{Buckley:2016kvr}
M.~R. Buckley, D.~Feld, S.~Macaluso, A.~Monteux, and D.~Shih, ``{Cornering
  Natural SUSY at LHC Run II and Beyond},''
  \href{http://dx.doi.org/10.1007/JHEP08(2017)115}{{\em JHEP} {\bf 08} (2017)
  115},
\href{http://arxiv.org/abs/1610.08059}{{\tt arXiv:1610.08059 [hep-ph]}}.
%%CITATION = ARXIV:1610.08059;%%.

\bibitem{SUSYSummaryPlot}
{ATLAS Collaboration}.
\newblock
  \url{https://atlas.web.cern.ch/Atlas/GROUPS/PHYSICS/CombinedSummaryPlots/SUSY/ATLAS_SUSY_Stop_tLSP/ATLAS_SUSY_Stop_tLSP.png}.

\bibitem{SUSY-2016-15}
{ATLAS Collaboration}, ``{Search for a scalar partner of the top quark in the
  jets plus missing transverse momentum final state at $\sqrt{s} =
  13\;\mbox{TeV}$ with the ATLAS detector},''
  \href{http://dx.doi.org/10.1007/JHEP12(2017)085}{{\em JHEP} {\bf 12} (2017)
  085}, \href{http://arxiv.org/abs/1709.04183}{{\tt arXiv:1709.04183
  [hep-ex]}}.

\bibitem{SUSY-2016-16}
{ATLAS Collaboration}, ``{Search for top-squark pair production in final states
  with one lepton, jets, and missing transverse momentum using
  $36\;\mbox{fb}^{-1}$ of $\sqrt{s} = 13\;\mbox{TeV}$ $pp$ collision data with
  the ATLAS detector},'' \href{http://arxiv.org/abs/1711.11520}{{\tt
  arXiv:1711.11520 [hep-ex]}}.

\bibitem{SUSY-2016-17}
{ATLAS Collaboration}, ``{Search for direct top squark pair production in final
  states with two leptons in $\sqrt{s} = 13\;\mbox{TeV}$ $pp$ collisions with
  the ATLAS detector},''
  \href{http://dx.doi.org/10.1140/epjc/s10052-017-5445-x}{{\em Eur. Phys. J. C}
  {\bf 77} (2017)  898}, \href{http://arxiv.org/abs/1708.03247}{{\tt
  arXiv:1708.03247 [hep-ex]}}.

\bibitem{TOPQ-2014-07}
{ATLAS Collaboration}, ``{Measurement of Spin Correlation in Top--Antitop Quark
  Events and Search for Top Squark Pair Production in $pp$ Collisions at
  $\sqrt{s} = 8\;\mbox{TeV}$ Using the ATLAS Detector},''
  \href{http://dx.doi.org/10.1103/PhysRevLett.114.142001}{{\em Phys. Rev.
  Lett.} {\bf 114} (2015)  142001}, \href{http://arxiv.org/abs/1412.4742}{{\tt
  arXiv:1412.4742 [hep-ex]}}.

\bibitem{TOPQ-2013-04}
{ATLAS Collaboration}, ``{Measurement of the $t\bar{t}$ production
  cross-section using $e\mu$ events with $b$-tagged jets in $pp$ collisions at
  $\sqrt{s} = 7$ and $8\;\mbox{TeV}$ with the ATLAS detector},''
  \href{http://dx.doi.org/10.1140/epjc/s10052-016-4501-2}{{\em Eur. Phys. J. C}
  {\bf 74} (2014)  3109}, \href{http://arxiv.org/abs/1406.5375}{{\tt
  arXiv:1406.5375 [hep-ex]}}.

\bibitem{Sirunyan:2017cwe}
{\bf CMS} Collaboration, A.~M. Sirunyan {\em et al.}, ``{Search for
  supersymmetry in multijet events with missing transverse momentum in
  proton-proton collisions at 13 TeV},''
  \href{http://dx.doi.org/10.1103/PhysRevD.96.032003}{{\em Phys. Rev.} {\bf
  D96} (2017) no.~3, 032003},
\href{http://arxiv.org/abs/1704.07781}{{\tt arXiv:1704.07781 [hep-ex]}}.
%%CITATION = ARXIV:1704.07781;%%.

\bibitem{Sirunyan:2017kqq}
{\bf CMS} Collaboration, A.~M. Sirunyan {\em et al.}, ``{Search for new
  phenomena with the $M_{\mathrm {T2}}$ variable in the all-hadronic final
  state produced in proton–proton collisions at $\sqrt{s} = 13$ $\,\text
  {TeV}$},'' \href{http://dx.doi.org/10.1140/epjc/s10052-017-5267-x}{{\em Eur.
  Phys. J.} {\bf C77} (2017) no.~10, 710},
\href{http://arxiv.org/abs/1705.04650}{{\tt arXiv:1705.04650 [hep-ex]}}.
%%CITATION = ARXIV:1705.04650;%%.

\bibitem{Sirunyan:2017fsj}
{\bf CMS} Collaboration, A.~M. Sirunyan {\em et al.}, ``{Search for
  Supersymmetry in $pp$ Collisions at $\sqrt{s}=13\text{ }\text{ }\mathrm{TeV}$
  in the Single-Lepton Final State Using the Sum of Masses of Large-Radius
  Jets},'' \href{http://dx.doi.org/10.1103/PhysRevLett.119.151802}{{\em Phys.
  Rev. Lett.} {\bf 119} (2017) no.~15, 151802},
\href{http://arxiv.org/abs/1705.04673}{{\tt arXiv:1705.04673 [hep-ex]}}.
%%CITATION = ARXIV:1705.04673;%%.

\bibitem{Sirunyan:2017mrs}
{\bf CMS} Collaboration, A.~M. Sirunyan {\em et al.}, ``{Search for
  supersymmetry in events with one lepton and multiple jets exploiting the
  angular correlation between the lepton and the missing transverse momentum in
  proton-proton collisions at $\sqrt{s} = $ 13 TeV},''
  \href{http://dx.doi.org/10.1016/j.physletb.2018.03.028}{{\em Phys. Lett.}
  {\bf B780} (2018)  384--409},
\href{http://arxiv.org/abs/1709.09814}{{\tt arXiv:1709.09814 [hep-ex]}}.
%%CITATION = ARXIV:1709.09814;%%.

\bibitem{Sirunyan:2017uyt}
{\bf CMS} Collaboration, A.~M. Sirunyan {\em et al.}, ``{Search for physics
  beyond the standard model in events with two leptons of same sign, missing
  transverse momentum, and jets in proton–proton collisions at $\sqrt{s} =
  13\,\text {TeV} $},''
  \href{http://dx.doi.org/10.1140/epjc/s10052-017-5079-z}{{\em Eur. Phys. J.}
  {\bf C77} (2017) no.~9, 578},
\href{http://arxiv.org/abs/1704.07323}{{\tt arXiv:1704.07323 [hep-ex]}}.
%%CITATION = ARXIV:1704.07323;%%.

\bibitem{Sirunyan:2017hvp}
{\bf CMS} Collaboration, A.~M. Sirunyan {\em et al.}, ``{Search for
  supersymmetry in events with at least three electrons or muons, jets, and
  missing transverse momentum in proton-proton collisions at $ \sqrt{s}=13 $
  TeV},'' \href{http://dx.doi.org/10.1007/JHEP02(2018)067}{{\em JHEP} {\bf 02}
  (2018)  067},
\href{http://arxiv.org/abs/1710.09154}{{\tt arXiv:1710.09154 [hep-ex]}}.
%%CITATION = ARXIV:1710.09154;%%.

\bibitem{Han:2012fw}
Z.~Han, A.~Katz, D.~Krohn, and M.~Reece, ``{(Light) Stop Signs},''
  \href{http://dx.doi.org/10.1007/JHEP08(2012)083}{{\em JHEP} {\bf 08} (2012)
  083},
\href{http://arxiv.org/abs/1205.5808}{{\tt arXiv:1205.5808 [hep-ph]}}.
%%CITATION = ARXIV:1205.5808;%%.

\bibitem{Alves:2012ft}
D.~S.~M. Alves, M.~R. Buckley, P.~J. Fox, J.~D. Lykken, and C.-T. Yu, ``{Stops
  and $\not E_T$: The shape of things to come},''
  \href{http://dx.doi.org/10.1103/PhysRevD.87.035016}{{\em Phys. Rev.} {\bf
  D87} (2013) no.~3, 035016},
\href{http://arxiv.org/abs/1205.5805}{{\tt arXiv:1205.5805 [hep-ph]}}.
%%CITATION = ARXIV:1205.5805;%%.

\bibitem{Eifert:2014kea}
T.~Eifert and B.~Nachman, ``{Sneaky light stop},''
  \href{http://dx.doi.org/10.1016/j.physletb.2015.02.039}{{\em Phys. Lett.}
  {\bf B743} (2015)  218--223},
\href{http://arxiv.org/abs/1410.7025}{{\tt arXiv:1410.7025 [hep-ph]}}.
%%CITATION = ARXIV:1410.7025;%%.

\bibitem{Czakon:2014fka}
M.~Czakon, A.~Mitov, M.~Papucci, J.~T. Ruderman, and A.~Weiler, ``{Closing the
  stop gap},'' \href{http://dx.doi.org/10.1103/PhysRevLett.113.201803}{{\em
  Phys. Rev. Lett.} {\bf 113} (2014) no.~20, 201803},
\href{http://arxiv.org/abs/1407.1043}{{\tt arXiv:1407.1043 [hep-ph]}}.
%%CITATION = ARXIV:1407.1043;%%.

\bibitem{Buckley:2014fqa}
M.~R. Buckley, T.~Plehn, and M.~J. Ramsey-Musolf, ``{Top squark with mass close
  to the top quark},'' \href{http://dx.doi.org/10.1103/PhysRevD.90.014046}{{\em
  Phys. Rev.} {\bf D90} (2014) no.~1, 014046},
\href{http://arxiv.org/abs/1403.2726}{{\tt arXiv:1403.2726 [hep-ph]}}.
%%CITATION = ARXIV:1403.2726;%%.

\bibitem{Fuks:2014lva}
B.~Fuks, P.~Richardson, and A.~Wilcock, ``{Studying the sensitivity of monotop
  probes to compressed supersymmetric scenarios at the LHC},''
  \href{http://dx.doi.org/10.1140/epjc/s10052-015-3530-6}{{\em Eur. Phys. J.}
  {\bf C75} (2015) no.~7, 308},
\href{http://arxiv.org/abs/1408.3634}{{\tt arXiv:1408.3634 [hep-ph]}}.
%%CITATION = ARXIV:1408.3634;%%.

\bibitem{Ferretti:2015dea}
G.~Ferretti, R.~Franceschini, C.~Petersson, and R.~Torre, ``{Spot the stop with
  a b-tag},'' \href{http://dx.doi.org/10.1103/PhysRevLett.114.201801}{{\em
  Phys. Rev. Lett.} {\bf 114} (2015)  201801},
\href{http://arxiv.org/abs/1502.01721}{{\tt arXiv:1502.01721 [hep-ph]}}.
%%CITATION = ARXIV:1502.01721;%%.

\bibitem{Kobakhidze:2015scd}
A.~Kobakhidze, N.~Liu, L.~Wu, J.~M. Yang, and M.~Zhang, ``{Closing up a light
  stop window in natural SUSY at LHC},''
  \href{http://dx.doi.org/10.1016/j.physletb.2016.02.003}{{\em Phys. Lett.}
  {\bf B755} (2016)  76--81},
\href{http://arxiv.org/abs/1511.02371}{{\tt arXiv:1511.02371 [hep-ph]}}.
%%CITATION = ARXIV:1511.02371;%%.

\bibitem{Alwall:2014hca}
J.~Alwall, R.~Frederix, S.~Frixione, V.~Hirschi, F.~Maltoni, O.~Mattelaer,
  H.~S. Shao, T.~Stelzer, P.~Torrielli, and M.~Zaro, ``{The automated
  computation of tree-level and next-to-leading order differential cross
  sections, and their matching to parton shower simulations},''
  \href{http://dx.doi.org/10.1007/JHEP07(2014)079}{{\em JHEP} {\bf 07} (2014)
  079},
\href{http://arxiv.org/abs/1405.0301}{{\tt arXiv:1405.0301 [hep-ph]}}.
%%CITATION = ARXIV:1405.0301;%%.

\bibitem{Ball:2014uwa}
{\bf NNPDF} Collaboration, R.~D. Ball {\em et al.}, ``{Parton distributions for
  the LHC Run II},'' \href{http://dx.doi.org/10.1007/JHEP04(2015)040}{{\em
  JHEP} {\bf 04} (2015)  040},
\href{http://arxiv.org/abs/1410.8849}{{\tt arXiv:1410.8849 [hep-ph]}}.
%%CITATION = ARXIV:1410.8849;%%.

\bibitem{Sjostrand:2014zea}
T.~Sjöstrand, S.~Ask, J.~R. Christiansen, R.~Corke, N.~Desai, P.~Ilten,
  S.~Mrenna, S.~Prestel, C.~O. Rasmussen, and P.~Z. Skands, ``{An Introduction
  to PYTHIA 8.2},'' \href{http://dx.doi.org/10.1016/j.cpc.2015.01.024}{{\em
  Comput. Phys. Commun.} {\bf 191} (2015)  159--177},
\href{http://arxiv.org/abs/1410.3012}{{\tt arXiv:1410.3012 [hep-ph]}}.
%%CITATION = ARXIV:1410.3012;%%.

\bibitem{Beenakker:2015rna}
W.~Beenakker, C.~Borschensky, M.~Kramer, A.~Kulesza, E.~Laenen, S.~Marzani, and
  J.~Rojo, ``{NLO+NLL squark and gluino production cross-sections with
  threshold-improved parton distributions},''
  \href{http://dx.doi.org/10.1140/epjc/s10052-016-3892-4}{{\em Eur. Phys. J.}
  {\bf C76} (2016) no.~2, 53},
\href{http://arxiv.org/abs/1510.00375}{{\tt arXiv:1510.00375 [hep-ph]}}.
%%CITATION = ARXIV:1510.00375;%%.

\bibitem{Beenakker:1996ed}
W.~Beenakker, R.~Hopker, and M.~Spira, ``{PROSPINO: A Program for the
  production of supersymmetric particles in next-to-leading order QCD},''
\href{http://arxiv.org/abs/hep-ph/9611232}{{\tt arXiv:hep-ph/9611232
  [hep-ph]}}.
%%CITATION = HEP-PH/9611232;%%.

\bibitem{Beenakker:1997ut}
W.~Beenakker, M.~Kramer, T.~Plehn, M.~Spira, and P.~M. Zerwas, ``{Stop
  production at hadron colliders},''
  \href{http://dx.doi.org/10.1016/S0550-3213(98)00014-5}{{\em Nucl. Phys.} {\bf
  B515} (1998)  3--14},
\href{http://arxiv.org/abs/hep-ph/9710451}{{\tt arXiv:hep-ph/9710451
  [hep-ph]}}.
%%CITATION = HEP-PH/9710451;%%.

\bibitem{deFavereau:2013fsa}
{\bf DELPHES 3} Collaboration, J.~de~Favereau, C.~Delaere, P.~Demin,
  A.~Giammanco, V.~Lemaître, A.~Mertens, and M.~Selvaggi, ``{DELPHES 3, A
  modular framework for fast simulation of a generic collider experiment},''
  \href{http://dx.doi.org/10.1007/JHEP02(2014)057}{{\em JHEP} {\bf 02} (2014)
  057},
\href{http://arxiv.org/abs/1307.6346}{{\tt arXiv:1307.6346 [hep-ex]}}.
%%CITATION = ARXIV:1307.6346;%%.

\bibitem{Cacciari:2005hq}
M.~Cacciari and G.~P. Salam, ``{Dispelling the $N^{3}$ myth for the $k_t$
  jet-finder},'' \href{http://dx.doi.org/10.1016/j.physletb.2006.08.037}{{\em
  Phys. Lett.} {\bf B641} (2006)  57--61},
\href{http://arxiv.org/abs/hep-ph/0512210}{{\tt arXiv:hep-ph/0512210
  [hep-ph]}}.
%%CITATION = HEP-PH/0512210;%%.

\bibitem{Cacciari:2008gp}
M.~Cacciari, G.~P. Salam, and G.~Soyez, ``{The Anti-k(t) jet clustering
  algorithm},'' \href{http://dx.doi.org/10.1088/1126-6708/2008/04/063}{{\em
  JHEP} {\bf 04} (2008)  063},
\href{http://arxiv.org/abs/0802.1189}{{\tt arXiv:0802.1189 [hep-ph]}}.
%%CITATION = ARXIV:0802.1189;%%.

\bibitem{Cacciari:2011ma}
M.~Cacciari, G.~P. Salam, and G.~Soyez, ``{FastJet User Manual},''
  \href{http://dx.doi.org/10.1140/epjc/s10052-012-1896-2}{{\em Eur. Phys. J.}
  {\bf C72} (2012)  1896},
\href{http://arxiv.org/abs/1111.6097}{{\tt arXiv:1111.6097 [hep-ph]}}.
%%CITATION = ARXIV:1111.6097;%%.

\bibitem{Artoisenet:2012st}
P.~Artoisenet, R.~Frederix, O.~Mattelaer, and R.~Rietkerk, ``{Automatic
  spin-entangled decays of heavy resonances in Monte Carlo simulations},''
  \href{http://dx.doi.org/10.1007/JHEP03(2013)015}{{\em JHEP} {\bf 03} (2013)
  015},
\href{http://arxiv.org/abs/1212.3460}{{\tt arXiv:1212.3460 [hep-ph]}}.
%%CITATION = ARXIV:1212.3460;%%.

\bibitem{Muhlleitner:2003vg}
M.~Muhlleitner, A.~Djouadi, and Y.~Mambrini, ``{SDECAY: A Fortran code for the
  decays of the supersymmetric particles in the MSSM},''
  \href{http://dx.doi.org/10.1016/j.cpc.2005.01.012}{{\em Comput. Phys.
  Commun.} {\bf 168} (2005)  46--70},
\href{http://arxiv.org/abs/hep-ph/0311167}{{\tt arXiv:hep-ph/0311167
  [hep-ph]}}.
%%CITATION = HEP-PH/0311167;%%.

\bibitem{Veltman:1963th}
M.~J.~G. Veltman, ``{Unitarity and causality in a renormalizable field theory
  with unstable particles},''
\href{http://dx.doi.org/10.1016/S0031-8914(63)80277-3}{{\em Physica} {\bf 29}
  (1963)  186--207}.
%%CITATION = PHYSA,29,186;%%.

\bibitem{Dicus:1984fu}
D.~A. Dicus, E.~C.~G. Sudarshan, and X.~Tata, ``{Factorization Theorem for
  Decaying Spinning Particles},''
\href{http://dx.doi.org/10.1016/0370-2693(85)91571-0}{{\em Phys. Lett.} {\bf
  154B} (1985)  79--85}.
%%CITATION = PHLTA,154B,79;%%.

\bibitem{Berdine:2007uv}
D.~Berdine, N.~Kauer, and D.~Rainwater, ``{Breakdown of the Narrow Width
  Approximation for New Physics},''
  \href{http://dx.doi.org/10.1103/PhysRevLett.99.111601}{{\em Phys. Rev. Lett.}
  {\bf 99} (2007)  111601},
\href{http://arxiv.org/abs/hep-ph/0703058}{{\tt arXiv:hep-ph/0703058
  [hep-ph]}}.
%%CITATION = HEP-PH/0703058;%%.

\bibitem{Kauer:2007zc}
N.~Kauer, ``{Narrow-width approximation limitations},''
  \href{http://dx.doi.org/10.1016/j.physletb.2007.04.036}{{\em Phys. Lett.}
  {\bf B649} (2007)  413--416},
\href{http://arxiv.org/abs/hep-ph/0703077}{{\tt arXiv:hep-ph/0703077
  [hep-ph]}}.
%%CITATION = HEP-PH/0703077;%%.

\bibitem{Uhlemann:2008pm}
C.~F. Uhlemann and N.~Kauer, ``{Narrow-width approximation accuracy},''
  \href{http://dx.doi.org/10.1016/j.nuclphysb.2009.01.022}{{\em Nucl. Phys.}
  {\bf B814} (2009)  195--211},
\href{http://arxiv.org/abs/0807.4112}{{\tt arXiv:0807.4112 [hep-ph]}}.
%%CITATION = ARXIV:0807.4112;%%.

\bibitem{Martin:2009iq}
A.~D. Martin, W.~J. Stirling, R.~S. Thorne, and G.~Watt, ``{Parton
  distributions for the LHC},''
  \href{http://dx.doi.org/10.1140/epjc/s10052-009-1072-5}{{\em Eur. Phys. J.}
  {\bf C63} (2009)  189--285},
\href{http://arxiv.org/abs/0901.0002}{{\tt arXiv:0901.0002 [hep-ph]}}.
%%CITATION = ARXIV:0901.0002;%%.

\bibitem{Martin:2009bu}
A.~D. Martin, W.~J. Stirling, R.~S. Thorne, and G.~Watt, ``{Uncertainties on
  alpha(S) in global PDF analyses and implications for predicted hadronic cross
  sections},'' \href{http://dx.doi.org/10.1140/epjc/s10052-009-1164-2}{{\em
  Eur. Phys. J.} {\bf C64} (2009)  653--680},
\href{http://arxiv.org/abs/0905.3531}{{\tt arXiv:0905.3531 [hep-ph]}}.
%%CITATION = ARXIV:0905.3531;%%.

\bibitem{Martin:2010db}
A.~D. Martin, W.~J. Stirling, R.~S. Thorne, and G.~Watt, ``{Heavy-quark mass
  dependence in global PDF analyses and 3- and 4-flavour parton
  distributions},''
  \href{http://dx.doi.org/10.1140/epjc/s10052-010-1462-8}{{\em Eur. Phys. J.}
  {\bf C70} (2010)  51--72},
\href{http://arxiv.org/abs/1007.2624}{{\tt arXiv:1007.2624 [hep-ph]}}.
%%CITATION = ARXIV:1007.2624;%%.

\bibitem{Beneke:2011mq}
M.~Beneke, P.~Falgari, S.~Klein, and C.~Schwinn, ``{Hadronic top-quark pair
  production with NNLL threshold resummation},''
  \href{http://dx.doi.org/10.1016/j.nuclphysb.2011.10.021}{{\em Nucl. Phys.}
  {\bf B855} (2012)  695--741},
\href{http://arxiv.org/abs/1109.1536}{{\tt arXiv:1109.1536 [hep-ph]}}.
%%CITATION = ARXIV:1109.1536;%%.

\bibitem{Cacciari:2011hy}
M.~Cacciari, M.~Czakon, M.~Mangano, A.~Mitov, and P.~Nason, ``{Top-pair
  production at hadron colliders with next-to-next-to-leading logarithmic
  soft-gluon resummation},''
  \href{http://dx.doi.org/10.1016/j.physletb.2012.03.013}{{\em Phys. Lett.}
  {\bf B710} (2012)  612--622},
\href{http://arxiv.org/abs/1111.5869}{{\tt arXiv:1111.5869 [hep-ph]}}.
%%CITATION = ARXIV:1111.5869;%%.

\bibitem{Czakon:2012zr}
M.~Czakon and A.~Mitov, ``{NNLO corrections to top-pair production at hadron
  colliders: the all-fermionic scattering channels},''
  \href{http://dx.doi.org/10.1007/JHEP12(2012)054}{{\em JHEP} {\bf 12} (2012)
  054},
\href{http://arxiv.org/abs/1207.0236}{{\tt arXiv:1207.0236 [hep-ph]}}.
%%CITATION = ARXIV:1207.0236;%%.

\bibitem{Czakon:2012pz}
M.~Czakon and A.~Mitov, ``{NNLO corrections to top pair production at hadron
  colliders: the quark-gluon reaction},''
  \href{http://dx.doi.org/10.1007/JHEP01(2013)080}{{\em JHEP} {\bf 01} (2013)
  080},
\href{http://arxiv.org/abs/1210.6832}{{\tt arXiv:1210.6832 [hep-ph]}}.
%%CITATION = ARXIV:1210.6832;%%.

\bibitem{Czakon:2013goa}
M.~Czakon, P.~Fiedler, and A.~Mitov, ``{Total Top-Quark Pair-Production Cross
  Section at Hadron Colliders Through $O(\alpha^{4}_{S})$},''
  \href{http://dx.doi.org/10.1103/PhysRevLett.110.252004}{{\em Phys. Rev.
  Lett.} {\bf 110} (2013)  252004},
\href{http://arxiv.org/abs/1303.6254}{{\tt arXiv:1303.6254 [hep-ph]}}.
%%CITATION = ARXIV:1303.6254;%%.

\bibitem{Czakon:2011xx}
M.~Czakon and A.~Mitov, ``{Top++: A Program for the Calculation of the Top-Pair
  Cross-Section at Hadron Colliders},''
  \href{http://dx.doi.org/10.1016/j.cpc.2014.06.021}{{\em Comput. Phys.
  Commun.} {\bf 185} (2014)  2930},
\href{http://arxiv.org/abs/1112.5675}{{\tt arXiv:1112.5675 [hep-ph]}}.
%%CITATION = ARXIV:1112.5675;%%.

\bibitem{SUSY-2014-07}
{ATLAS Collaboration}, ``{ATLAS Run 1 searches for direct pair production of
  third-generation squarks at the Large Hadron Collider},''
  \href{http://dx.doi.org/10.1140/epjc/s10052-015-3726-9}{{\em Eur. Phys. J. C}
  {\bf 75} (2015)  510}, \href{http://arxiv.org/abs/1506.08616}{{\tt
  arXiv:1506.08616 [hep-ex]}}.

\bibitem{Cowan:2010js}
G.~Cowan, K.~Cranmer, E.~Gross, and O.~Vitells, ``{Asymptotic formulae for
  likelihood-based tests of new physics},''
  \href{http://dx.doi.org/10.1140/epjc/s10052-011-1554-0,
  10.1140/epjc/s10052-013-2501-z}{{\em Eur. Phys. J.} {\bf C71} (2011)  1554},
  \href{http://arxiv.org/abs/1007.1727}{{\tt arXiv:1007.1727
  [physics.data-an]}}.
[Erratum: Eur. Phys. J.C73,2501(2013)].
%%CITATION = ARXIV:1007.1727;%%.

\bibitem{ATLAS:2011tau}
{\bf The ATLAS Collaboration, The CMS Collaboration, The LHC Higgs Combination
  Group} Collaboration, G.~Aad {\em et al.}, ``{Procedure for the LHC Higgs
  boson search combination in Summer 2011},'' Tech. Rep. CMS-NOTE-2011-005.
  ATL-PHYS-PUB-2011-11, CERN, Geneva, Aug, 2011.
\newblock \url{http://cds.cern.ch/record/1379837}.

\bibitem{Battaglieri:2017aum}
M.~Battaglieri {\em et al.}, ``{US Cosmic Visions: New Ideas in Dark Matter
  2017: Community Report},''
\href{http://arxiv.org/abs/1707.04591}{{\tt arXiv:1707.04591 [hep-ph]}}.
%%CITATION = ARXIV:1707.04591;%%.

\bibitem{Martin:2008sv}
S.~P. Martin, ``{Diphoton decays of stoponium at the Large Hadron Collider},''
  \href{http://dx.doi.org/10.1103/PhysRevD.77.075002}{{\em Phys. Rev.} {\bf
  D77} (2008)  075002},
\href{http://arxiv.org/abs/0801.0237}{{\tt arXiv:0801.0237 [hep-ph]}}.
%%CITATION = ARXIV:0801.0237;%%.

\bibitem{Martin:2009dj}
S.~P. Martin and J.~E. Younkin, ``{Radiative corrections to stoponium
  annihilation decays},''
  \href{http://dx.doi.org/10.1103/PhysRevD.80.035026}{{\em Phys. Rev.} {\bf
  D80} (2009)  035026},
\href{http://arxiv.org/abs/0901.4318}{{\tt arXiv:0901.4318 [hep-ph]}}.
%%CITATION = ARXIV:0901.4318;%%.

\bibitem{Younkin:2009zn}
J.~E. Younkin and S.~P. Martin, ``{QCD corrections to stoponium production at
  hadron colliders},'' \href{http://dx.doi.org/10.1103/PhysRevD.81.055006}{{\em
  Phys. Rev.} {\bf D81} (2010)  055006},
\href{http://arxiv.org/abs/0912.4813}{{\tt arXiv:0912.4813 [hep-ph]}}.
%%CITATION = ARXIV:0912.4813;%%.

\bibitem{Kumar:2014bca}
N.~Kumar and S.~P. Martin, ``{LHC search for di-Higgs decays of stoponium and
  other scalars in events with two photons and two bottom jets},''
  \href{http://dx.doi.org/10.1103/PhysRevD.90.055007}{{\em Phys. Rev.} {\bf
  D90} (2014) no.~5, 055007},
\href{http://arxiv.org/abs/1404.0996}{{\tt arXiv:1404.0996 [hep-ph]}}.
%%CITATION = ARXIV:1404.0996;%%.

\bibitem{Batell:2015zla}
B.~Batell and S.~Jung, ``{Probing Light Stops with Stoponium},''
  \href{http://dx.doi.org/10.1007/JHEP07(2015)061}{{\em JHEP} {\bf 07} (2015)
  061},
\href{http://arxiv.org/abs/1504.01740}{{\tt arXiv:1504.01740 [hep-ph]}}.
%%CITATION = ARXIV:1504.01740;%%.

\end{thebibliography}\endgroup

\end{document}